\documentclass[11pt,a4paper]{emulateapj}
\usepackage{natbib}
\usepackage{mathtools}
\usepackage{graphicx}
\usepackage{aas_macros}
\newcommand{\unit}[1]{\ensuremath{\, \mathrm{#1}}}

\usepackage[usenames,dvipsnames]{xcolor} 

\submitted{To appear in \emph{The Astrophysical Journal}}
\slugcomment{To appear in The Astrophysical Journal}

\shorttitle{Effect of low-T opacity on low-Z AGB models} 

\shortauthors{Constantino et al.}

\providecommand{\hst}{\hspace{-0.2cm}}

\begin{document}

\title{On the necessity of composition-dependent low-temperature opacity\\ in metal-poor AGB stars} 

\author{Thomas Constantino\altaffilmark{1}, Simon Campbell\altaffilmark{1}, Pilar Gil-Pons \altaffilmark{2} and John Lattanzio\altaffilmark{1}}
\affil{\altaffilmark{1}Monash Centre for Astrophysics (MoCA), School of Mathematical Sciences,  Monash University, Victoria, 3800, Australia\\
\altaffilmark{2}Department of Applied Physics, Polytechnic University of Catalonia, 08860 Barcelona, Spain}
\email{Thomas.Constantino@monash.edu}

\begin{abstract} 
The vital importance of composition-dependent low-temperature opacity in low-mass ($M \leq 3$\,M$_\odot$) asymptotic giant branch (AGB) stellar models of metallicity $Z\geq 0.001$ has recently been demonstrated \citep[e.g.][]{2002A&A...387..507M,2010MNRAS.408.2476V}.  Its significance to more metal-poor, intermediate mass ($M \geq 2.5$\,M$_\odot$) models has yet to be investigated.  We show that its inclusion in lower-metallicity models ([Fe/H] $\leq -2$) is essential, and that there exists no threshold metallicity below which composition-dependent molecular opacity may be neglected.  We find it to be crucial in all intermediate-mass models investigated ([Fe/H] $\leq -2$ and $2.5 \leq M/\text{M}_\odot \leq 5$), because of the evolution of the surface chemistry, including the orders of magnitude increase in the abundance of molecule-forming species.  Its effect on these models mirrors that previously reported for higher-metallicity models -- increase in radius, decrease in $T_\text{eff}$, faster mass loss, shorter thermally pulsing AGB lifetime, reduced enrichment in third dredge-up products (by a factor of three to ten), and an increase in the mass limit for hot bottom burning.  We show that the evolution of low-metallicity models with composition-dependent low-temperature opacity is relatively independent of initial metal abundance because its contribution to the opacity is far outweighed by changes due to dredge-up.  Our results imply a significant reduction in the expected number of nitrogen-enhanced metal-poor stars, which may help explain their observed paucity.  We note that these findings are partially a product of the macrophysics adopted in our models, in particular the \citet{1993ApJ...413..641V} mass loss rate which is strongly dependent on radius.
\end{abstract}

\keywords{opacity --- stars: AGB and post-AGB --- stars: evolution --- stars: interiors --- stars: Population III}

\section{Introduction}
The asymptotic giant branch (AGB) is the final phase of nuclear burning experienced by stars with initial mass of roughly $0.8-8$\,M$_\odot$.  These stars are composed of a large convective envelope above hydrogen burning and helium burning shells that surround a degenerate CO core.  Late in this phase the envelope is ejected in a stellar wind, enriching the interstellar medium with nuclearly processed material.  Thermally pulsing (TP) AGB stars are thought to be producers of the s-process elements \citep{1999ARA&A..37..239B}, which are brought to the surface, along with carbon from helium burning, during episodes of the third dredge-up (3DU).  Since \citet{1975ApJ...196..805S}, models of higher mass AGB stars have shown nitrogen production via the process now referred to as hot bottom burning (HBB) in which the convective envelope penetrates into regions where the CNO-cycle is operating.  This finding was later supported by observations of AGB stars in the Magellanic clouds \citep{2007MNRAS.378.1089M}.   The resultant material has long been a prime suspect for the characteristic abundance patterns in globular cluster (GC) stars \citep{1981ApJ...245L..79C,2004ARA&A..42..385G}.  AGB material is also thought to contribute to the formation of carbon-enhanced metal-poor (CEMP) and nitrogen-enhanced metal-poor (NEMP) stars \citep[e.g.][]{2011MNRAS.412..843S,2012A&A...547A..76P}. 

When the outer region of a star is cool enough ($T \lesssim 5000 \unit{K}$), molecules make an important contribution to the opacity.  During the AGB phase, the third dredge-up and HBB can radically change the chemistry of the surface, but stellar evolution codes have historically used low-T Rosseland mean opacity data that assume a scaled-solar abundance pattern \citep[e.g.][]{1975ApJS...29..363A,1994ApJ...437..879A,2005ApJ...623..585F}.  There have been, however, a few notable exceptions.  By using a polynomial fit for the opacity due to CN molecules, \citet{1975ApJ...200..682S} were the first to run models that accounted for the changing composition during the AGB phase.  \citet{1983ApJ...272..773A} then produced tables for three additional C/O ratios (0.95, 1.0 and 1.05).  \citet{1986A&A...154..267L} calculated the opacity for a $\text{C/O} = 2$ mixture for C-rich AGB models.  \citet{1989A&AS...77....1B} used the \citet{1983ApJ...272..773A} tables to produce a fit for the opacity as a function of carbon abundance.  More recently, \citet{2002A&A...387..507M} demonstrated the shortcomings of using scaled-solar abundance low-T opacity in synthetic AGB models, by computing the opacity for mixtures of a range of molecular species for which data were available.  These effects were also apparent when composition-dependent molecular opacity was later used in full stellar models \citep{2007ApJ...667..489C,2008ASPC..388..183K,2009A&A...508.1343W}.  Sets of low-temperature opacity tables for enhancements in carbon and nitrogen abundance using more detailed calculations have since been made available \citep{2009A&A...494..403L}, as well as the online {\sc aesopus} tool which can generate low-T opacity tables for any abundance mixture with data for about 300 atomic and 500 molecular species \citep{2009A&A...508.1539M}.  

There are a number of important consequences from following the changing molecular composition in low to intermediate-mass models.  When C/O first exceeds unity, because of 3DU, the opacity in the outer envelope increases, the surface cools, and the mass loss rate increases.  This reduces the AGB lifetime, halts HBB in certain cases \citep{2007A&A...467.1139M,2010MNRAS.408.2476V}, and reduces chemical yields of certain species \citep[e.g.][]{2002A&A...387..507M}.  The final envelope C/O ratio in such models is more consistent with observations \citep{2002A&A...387..507M}.  

Studies into the effect of composition-dependent low-T opacity have hitherto been confined to higher metallicity stars ($Z \ge 0.001$), apart from \citet{2007ApJ...667..489C} in which a 2\,M$_\odot$, $Z=10^{-4}$ model was examined.  Since the molecular composition (and therefore opacity) is most sensitive to element abundances when $\text{C/O} \sim 1$ (by number), very little effect has been seen unless there is enough dredge-up to reach C/O $> 1$ \citep{2010MNRAS.408.2476V}.  Low-metallicity stars can more easily become carbon-rich because there is less oxygen in the envelope for the dredged-up carbon to overcome.  They can also experience orders of magnitude increase in total C+N+O which leads to an opacity increase irrespective of C/O.  Despite this, it has been argued that the adoption of composition-dependent low-temperature opacity in low-metallicity models may not be important because of the lower absolute abundance of molecules and higher temperature, which inhibits molecule formation \citep{2002A&A...387..507M,2007A&A...467.1139M}.  The single low-mass model from \citet{2007ApJ...667..489C}, however, showed a halving of the AGB lifetime by taking into account C and N enhancement.  This disagreement highlights the need for the contribution of composition-dependent low-T opacity to low-metallicity models to be tested over a range of initial mass and metallicity with a suite of full stellar models.  This work forms the basis of this study.

Abundance anticorrelations, such as Na-O, are ubiquitous in Galactic globular clusters \citep{2010A&A...516A..55C}, and pollution from an earlier generation of higher-mass AGB stars may be responsible for them \citep[e.g.][]{2013MNRAS.tmp.2760V}.  AGB stars can produce these anticorrelations via hot bottom burning where the CNO and Ne-Na cycles are active \citep{2004ARA&A..42..385G}.  This hypothesis constrains chemical yields because there is a limited internal spread of C+N+O in clusters.  For instance, in M13 the total C+N+O content differs between populations by roughly a factor of two \citep{2005AJ....129..303C} and it is constant to within observational errors in M4 \citep{1999AJ....118.1273I}.  Unfortunately this evidence is often equivocal.  Models of a population with a factor of three difference in C+N+O in NGC 1851 best reproduce the observed split in the subgiant branch \citep{2009MNRAS.399..934V} and \citet{2009ApJ...695L..62Y} observed a spread in C+N+O of a factor of 4 in four RGB stars in the same cluster.  By contrast, also in NGC 1851, \citet{2010ApJ...722L..18V} found no evidence for a spread in 15 RGB stars and \citet{2012A&A...539A..19G} put an upper limit of $\Delta$[(C+N+O)/Fe] $\leq 0.2$ by observing blue horizontal branch stars.  In higher-metallicity low-mass models, composition-dependent low-T opacity has been shown to reduce the enrichment of the envelope in C+N+O by truncating the evolution, but so far only in masses too low for HBB \citep{2009MNRAS.399L..54V}.  Whether or not this distinction based on initial mass holds for lower-metallicity models is important for the GC self-enrichment scenario.

Various estimates put the fraction of extremely metal-poor (EMP) stars ([Fe/H] $<-2$) that are also carbon-enhanced ([C/Fe] $\geq 1$) between around 9\% and greater than 20\% \citep{2006ApJ...652.1585F,2006ApJ...652L..37L,2012ApJ...744..195C}.  This fraction increases with decreasing metallicity \citep{2005AJ....130.2804R} and with increasing distance from the Galactic plane \citep{2006ApJ...652.1585F}.  Observations suggest that the s-process rich (CEMP-s) stars are all binaries \citep{2005ApJ...625..825L}, supporting the hypothesis that mass transfer from an AGB companion is responsible for the carbon-enhancement in these stars \citep[for models see e.g.][]{2008MNRAS.389.1828S}.  There is a paucity of observed nitrogen-enhanced metal-poor (NEMP) stars ($\text{[N/Fe]} > 1$ and [N/C] $> 0.5$; \citealt{2007ApJ...658.1203J}), with a handful of exceptions at very low metallicity $\text{[Fe/H]} < -2.8$ \citep{2010A&A...509A..93M}.  In population synthesis models \citep[e.g.][]{2009A&A...508.1359I,2012A&A...547A..76P,2013MNRAS.432L..46S} the CEMP/NEMP ratio is sensitive to the mass threshold above which HBB occurs.  This mass limit differs considerably from code to code and at higher metallicity has been shown to be significantly increased by adopting composition-dependent low-T opacity \citep{2007A&A...467.1139M}.  If this is also true for low-metallicity models it would reduce the predicted number of NEMP stars.  Another important ingredient that determines the HBB limit, and a poorly constrained one at low-metallicity, is the mass loss formulation used, which can differ by orders of magnitude and therefore affect lifetimes enough to control whether or not HBB converts C to N.  Composition-dependent low-T opacity rapidly affects the radius, and later the luminosity, which can both alter the mass loss rate, depending on the formulation, making code comparisons difficult.  Modeling a population of binary systems and mass transfer adds yet another layer of complexity.  

In recent years the availability of molecular opacity data has for the first time allowed stellar evolution codes to accurately account for the significant composition changes that can occur during the AGB phase.  The composition changes are most extreme for low-metallicity models but the consequences have only been studied in detail for the more metal-rich regime ($Z \ge 0.001$). Here we extend this inquiry to low metallicity.  We quantify the effects and establish their mass and metallicity dependence.  We also consider how any changes in the evolution affect CEMP and NEMP formation and the role of AGB stars in the globular cluster self-pollution scenario.

\section{Methods}
\subsection{Stellar structure code}
We use the 1-dimensional Monash University stellar structure code {\sc monstar} which has been described in detail previously \citep[e.g.][]{1996ApJ...473..383F,2008A&A...490..769C}.  We have since updated the code to use the OPAL \citep{2002ApJ...576.1064R}, Helmholtz \citep{2000ApJS..126..501T} and Timmes equations of state \citep{1999ApJS..125..277T}.  In this study we use Helmholtz EOS for $T > 1 \unit{MK}$, OPAL EOS for $T <2 \unit{MK}$, and a linear blend of the two in the overlapping region.  During the AGB phase we use instantaneous mixing of chemical species and determine convective boundaries using a ``search for convective neutrality'' \citep{1986ApJ...311..708L}.  During core helium burning we have no overshoot.  This eliminates the stochastic nature of the mixing during this phase \citep{1985ApJ...296..204C}, ensuring that each model of a given mass and metallicity begins the AGB with the same H-exhausted core mass (so that opacity is the only variable).  The mass loss for the RGB is \citet{1975MSRSL...8..369R} with $\eta_\text{R} = 0.4$ and for the AGB it is the \citet{1993ApJ...413..641V} rate.  

\subsection{Updated low-T opacity treatment}
For this study we implement a custom grid of {\sc aesopus} low-temperature opacity tables with three values for the abundance of H, 13 for C, 6 for N, 38 for C/O, and 12 for metallicity $Z$ (Table \ref{table_opacs}).  We also have a grid that is used for models that are initially metal-free.  It has the same mass fraction C, N, and O enhancement as the $Z=10^{-6}$ set, which we later show is reasonable because the $Z=0$ and $\text{[Fe/H]}=-4$ models evolve to have similar C, N, and O surface abundances (Table \ref{table_model_summary}).  We account for $-1.5 \le$[C/O]$\le +2.5$ and enhancements up to [C/Fe] = [N/Fe] $= +4$.  We have tables for 38 different C/O values in order to properly resolve the opacity near and just below C/O $=1$, where because of the strength of the molecular bonds in CO and SiO it rapidly changes (see Figure 16 in \citealt{2009A&A...508.1539M}).  Our resolution is considerably finer than others in the literature \citep[e.g.][]{2010MNRAS.408.2476V,2014MNRAS.tmp....3F}, although we note that \citet{2013MNRAS.434..488M} have very recently fully integrated {\sc aesopus} into their synthetic AGB code so that the opacity for the particular composition in the envelope during the evolution is computed directly.  The increase in the number of tables in our code (to almost 116,000) only requires additional memory.  We minimize interpolation errors by using linear interpolation in C/O, [C/Fe], [N/Fe], hydrogen $X$, $T$, and $R=\rho/T^3_6$ (where $T_6$ is $T$ in MK), and logarithmic interpolation in $Z$.  An important improvement in {\sc aesopus} compared to the earlier \citet{2009A&A...494..403L} tables is that it can account for an increase in envelope oxygen abundance, which can be considerable in low-metallicity models, and is obviously essential for determining if C/O $>1$.  We hereafter refer to composition-independent low-T opacity as the ``old'' opacity and composition-dependent low-T opacity as ``new'' opacity.  Both of these use the {\sc aesopus} tables except that in the ``old'' case only the initial abundance is used to calculate low-T opacity.  The composition-independent opacity models serve as a control group.

\subsection{Stellar models}
We have computed the evolution of 11 pairs of stellar models (with and without the new opacity) from the pre-main sequence to the late AGB.  The grid of models is shown in Table \ref{table_model_summary}.  We use a scaled-solar abundance from \citet{2009ARA&A..47..481A} with oxygen-enhancement [O/Fe] = +0.4 to mimic the effect of an alpha-enhancement abundance pattern (in all but the $\text{[Fe/H]}=-1$ model), and an initial helium mass fraction of $Y = 0.245$.  The models have metallicity [Fe/H] $\leq -1$, which is \citep[with the exception of one model in][]{2007ApJ...667..489C} more metal-poor than has been studied previously.  Composition-dependent low-T opacity has also been recently used in other codes for low-metallicity models but its effect has not been specifically analyzed \citep[e.g.][]{2009ApJ...696..797C,2012ApJ...747....2L}.  The uptake is so far not ubiquitous, it is not yet included in MESA for instance \citep{2013ApJS..208....4P}.  Our metallicity range is relevant to CEMP (and NEMP) stars and globular cluster chemical evolution.  The models span a mass range from 1.25 to 5\,M$_\odot$ which covers models that have a mass much too low for HBB to those which experience strong HBB.  This mass range was chosen so that we could investigate how the mass limit for HBB is affected, complementing work already done for higher-metallicity models \citep[e.g.][]{2009MNRAS.399L..54V}.  We examined 2.5\,M$_\odot$ models down to $Z=0$ to see if there is a metallicity cut-off below which HBB is unaffected by the updated opacity treatment.  We also tested the sensitivity of our results to a change in the mixing length parameter in the mixing length theory (MLT).  In the interest of time we did not attempt to restart several of the models with the old opacity after convergence problems, if the effect on the evolution was already obvious.  Therefore, some of these models still had varying amounts of the envelope remaining at the end of computation.

\section{Stellar model results}

\begin{deluxetable}{ccccccc}
\tabletypesize{\scriptsize}
  \tablecaption{Composition and parameters for the low-T opacity tables used in this study. \label{table_opacs}}
\tablewidth{0pt}
\tablehead{
 \colhead{$Z$} & \colhead{$X$} & \colhead{[C/Fe]} & \colhead{[N/Fe]} & \colhead{C/O} & \colhead{$\log{R}$} & \colhead{$\log{T}$}
}
\startdata
0 & 0.50 & -3.0 & -3.0 & 0.017 & -8.0 & 3.20 \\
$1 \times 10^{-6}$ & 0.65 & -1.5 &  0.0 & 0.054 & -7.5 & 3.22 \\
$3 \times 10^{-6}$ & 0.80 & -1.0 &  +1.0 & 0.170 & -7.0 & 3.24 \\
$1 \times 10^{-5}$ &\nodata & -0.5 &  +2.0 & 0.380 & -6.5 & 3.26 \\
$3 \times 10^{-5}$ &\nodata &  0.0 &  +3.0 & 0.537 & -6.0 & 3.28 \\
$1 \times 10^{-4}$ &\nodata &  +0.5 &  +4.0 & 0.708 & -5.5 & 3.30 \\
$3 \times 10^{-4}$ &\nodata &  +1.0 &\nodata & 0.813 & -5.0 & 3.32 \\
0.001 &\nodata &  +1.5 &\nodata & 0.852 & -4.5 & 3.34 \\
0.0025 &\nodata &  +2.0 &\nodata & 0.872 & -4.0 & 3.36 \\
0.005 &\nodata &  +2.5 &\nodata & 0.882 & -3.5 & 3.38 \\
0.01 &\nodata &  +3.0 &\nodata & 0.892 & -3.0 & 3.40 \\
0.02 &\nodata &  +3.5 &\nodata & 0.902 & -2.5 & 3.42 \\
0.04 &\nodata &  +4.0 &\nodata & 0.913 & -2.0 & 3.44 \\
\nodata &\nodata & &\nodata & 0.923 & -1.5 & 3.46 \\
\nodata &\nodata & &\nodata & 0.934 & -1.0 & 3.48 \\
\nodata &\nodata & &\nodata & 0.945 & -0.5 & 3.50 \\
\nodata &\nodata & &\nodata & 0.956 &  0.0 & 3.52 \\
\nodata &\nodata & &\nodata & 0.967 &  +0.5 & 3.54 \\
\nodata &\nodata & &\nodata & 0.978 &  +1.0 & 3.56 \\
\nodata &\nodata & &\nodata & 0.987 &\nodata & 3.58 \\
\nodata &\nodata & &\nodata & 0.994 &\nodata & 3.60 \\
\nodata &\nodata & &\nodata & 0.998 &\nodata & 3.62 \\
\nodata &\nodata & &\nodata & 1.000 &\nodata & 3.64 \\
\nodata &\nodata & &\nodata & 1.003 &\nodata & 3.66 \\
\nodata &\nodata & &\nodata & 1.008 &\nodata & 3.68 \\
\nodata &\nodata & &\nodata & 1.015 &\nodata & 3.70 \\
\nodata &\nodata & &\nodata & 1.024 &\nodata & 3.75 \\
\nodata &\nodata & &\nodata & 1.048 &\nodata & 3.80 \\
\nodata &\nodata & &\nodata & 1.072 &\nodata & 3.85 \\
\nodata &\nodata & &\nodata & 1.123 &\nodata & 3.90 \\
\nodata &\nodata & &\nodata & 1.203 &\nodata & 3.95 \\
\nodata &\nodata & &\nodata & 1.350 &\nodata & 4.00 \\
\nodata &\nodata & &\nodata & 1.699 &\nodata & 4.05 \\
\nodata &\nodata & &\nodata & 2.401 &\nodata &\nodata \\
\nodata &\nodata & &\nodata & 5.374 &\nodata &\nodata \\
\nodata &\nodata & &\nodata & 16.995 &\nodata &\nodata \\
\nodata &\nodata & &\nodata & 53.743 &\nodata &\nodata \\
\nodata &\nodata & &\nodata & 169.949 &\nodata &\nodata \\
\enddata
\vspace{-0.20cm}
\tablecomments{$Z$ refers to the total metal content before C, N, and O alteration and $\log{R}=\log{\rho/T_6^3}$ where $T_6$ is $T$ in MK.  The tables were generated online with {\sc aesopus} using the \citet{2007SSRv..130..105G} solar abundance.}
\end{deluxetable}

\placetable{table_opacs}

\begin{deluxetable*}{cccccccccccccccccc}
\tabletypesize{\footnotesize}
\tablecolumns{18}

   \tablecaption{Summary of the properties of each model.  \label{table_model_summary}}
\tablehead{
 \colhead{$M$} & 
 \colhead{\hst [Fe/H]} &
 \colhead{\hst $\kappa$} &
 \colhead{\hst $\alpha_\text{MLT}$} &
 \colhead{\hst $n_\text{TP}$} &
 \colhead{\hst $n_\text{3DU}$} & 
 \colhead{\hst $t_\text{AGB}$} &
 \colhead{\hst $\lambda_\text{peak}$} & 
 \colhead{\hst $M_\text{core,f}$} &
 \colhead{\hst $M_\text{f}$} & 
 \colhead{\hst $M_\text{env,f}$} &
 \colhead{\hst $T_\text{bce,max}$ }& 
 \colhead{\hst $Y$} &
 \colhead{\hst [C/Fe]} &
 \colhead{\hst [N/Fe]} &
 \colhead{\hst [O/Fe]} & 
 \colhead{\hst R$_\text{CNO}$} &
 \colhead{\hst C/O$_\text{f}$}
}

\startdata
1.25&\hst$-2$ &\hst$\kappa_0$ &\hst1.6 &\hst35        &\hst1   &\hst3.05 &\hst0.18 &\hst0.766 &\hst0.839 &\hst0.073 &\hst2.67 &\hst0.269 &\hst1.41 &\hst0.62 &\hst0.45 &\hst4  &\hst3.3  \\
1.25&\hst$-2$ &\hst$\kappa_X$ &\hst1.6 &\hst31        &\hst1   &\hst2.89 &\hst0.17 &\hst0.750 &\hst0.750 &\hst0.000 &\hst2.17 &\hst0.269 &\hst1.48 &\hst0.58 &\hst0.45 &\hst4  &\hst3.2  \\
1.75&\hst$-2$ &\hst$\kappa_0$ &\hst1.6 &\hst106       &\hst92  &\hst3.10 &\hst0.62 &\hst0.843 &\hst0.954 &\hst0.111 &\hst35.7 &\hst0.369 &\hst3.25 &\hst2.93 &\hst1.15 &\hst220 &\hst39 \\
1.75&\hst$-2$ &\hst$\kappa_X$ &\hst1.6 &\hst25        &\hst12  &\hst1.90 &\hst0.26 &\hst0.726 &\hst0.726 &\hst0.000 &\hst4.01 &\hst0.278 &\hst2.23 &\hst1.04 &\hst0.61 &\hst20 &\hst13 \\
2.5 &\hst$-2$ &\hst$\kappa_0$ &\hst1.6 &\hst197       &\hst195 &\hst2.42 &\hst0.91 &\hst0.877 &\hst1.977 &\hst1.100 &\hst72.6 &\hst0.411 &\hst1.97 &\hst4.04 &\hst1.36 &\hst420  &\hst1.8  \\
2.5 &\hst$-2$ &\hst$\kappa_X$ &\hst1.6 &\hst19        &\hst17  &\hst0.71 &\hst0.83 &\hst0.745 &\hst1.335 &\hst0.590 &\hst12.3 &\hst0.277 &\hst2.57 &\hst1.53 &\hst0.66 &\hst42   &\hst26  \\
2.5 &\hst$-3$ &\hst$\kappa_0$ &\hst1.6 &\hst240       &\hst238 &\hst 2.20&\hst0.92 &\hst0.910 &\hst2.374 &\hst1.464 &\hst78.8 &\hst0.442 &\hst2.82 &\hst5.04 &\hst2.17 &\hst4100 &\hst1.4  \\
2.5 &\hst$-3$ &\hst$\kappa_X$ &\hst1.6 &\hst19        &\hst17  &\hst0.63 &\hst0.86 &\hst0.755 &\hst1.471 &\hst0.716 &\hst15.6 &\hst0.286 &\hst3.61 &\hst2.56 &\hst1.41 &\hst460  &\hst49  \\
2.5 &\hst$-4$ &\hst$\kappa_0$ &\hst1.6 &\hst193       &\hst191 &\hst 1.81&\hst0.93 &\hst0.882 &\hst2.414 &\hst1.532 &\hst78.1 &\hst0.435 &\hst3.78 &\hst6.00 &\hst3.12 &\hst37000&\hst1.4  \\
2.5 &\hst$-4$ &\hst$\kappa_X$ &\hst1.6 &\hst22        &\hst20  &\hst0.57 &\hst0.89 &\hst0.775 &\hst1.340 &\hst0.565 &\hst21.1 &\hst0.300 &\hst4.65 &\hst3.64 &\hst2.39 &\hst4900 &\hst55  \\
2.5 &\hst$-\infty$&\hst$\kappa_0$ &\hst1.6&\hst206   &\hst205 &\hst1.75&\hst0.94 &\hst0.875&\hst0.065 &\hst2.462 &\hst78.5  &\hst0.478&\hst3.79  &\hst5.97 &\hst3.08 &\hst\nodata  &\hst1.6 \\ 
2.5 &\hst$-\infty$&\hst$\kappa_X$ &\hst1.6&\hst26    &\hst25  &\hst0.64 &\hst0.92 &\hst0.748 &\hst0.172 &\hst1.438 &\hst19.7 &\hst0.369 &\hst4.78 &\hst3.94 &\hst2.49 &\hst\nodata &\hst60 \\ 
3 &\hst$-2$ &\hst$\kappa_0$ &\hst1.6 &\hst232         &\hst224 &\hst1.89 &\hst0.99 &\hst0.897 &\hst1.234 &\hst0.337 &\hst77.0 &\hst0.415 &\hst1.99 &\hst4.02 &\hst1.23 &\hst400 &\hst2.0  \\
3 &\hst$-2$ &\hst$\kappa_X$ &\hst1.6 &\hst25          &\hst22  &\hst0.44 &\hst0.95 &\hst0.816 &\hst1.677 &\hst0.861 &\hst34.4 &\hst0.274 &\hst2.55 &\hst1.42 &\hst0.66 &\hst41  &\hst25  \\
4 &\hst$-2$ &\hst$\kappa_0$ &\hst1.6 &\hst295         &\hst292 &\hst1.43 &\hst0.95 &\hst0.921 &\hst2.144 &\hst1.223 &\hst86.2 &\hst0.430 &\hst1.82 &\hst3.89 &\hst1.07 &\hst290 &\hst1.9  \\
4 &\hst$-2$ &\hst$\kappa_X$ &\hst1.6 &\hst85          &\hst81  &\hst0.65 &\hst0.95 &\hst0.884 &\hst1.991 &\hst1.107 &\hst82.9 &\hst0.328 &\hst1.64 &\hst3.39 &\hst0.74 &\hst96  &\hst3.2  \\
4 &\hst$-3$ &\hst$\kappa_0$ &\hst1.6 &\hst309         &\hst306 &\hst1.38 &\hst0.95 &\hst0.925 &\hst3.844 &\hst2.919 &\hst87.8 &\hst0.459 &\hst2.71 &\hst4.88 &\hst1.99 &\hst2800&\hst1.6  \\
4 &\hst$-3$ &\hst$\kappa_X$ &\hst1.6 &\hst103         &\hst100 &\hst0.77 &\hst0.95 &\hst0.882 &\hst2.174 &\hst1.292 &\hst84.3 &\hst0.344 &\hst2.68 &\hst4.49 &\hst1.69 &\hst1200&\hst3.9  \\
5 &\hst$-1$ &\hst$\kappa_0$ &\hst1.6 &\hst142         &\hst138 &\hst0.69 &\hst0.93 &\hst0.924 &\hst1.915 &\hst0.991 &\hst88.8 &\hst0.354 &\hst0.61 &\hst2.42 &\hst-0.15&\hst21  &\hst3.0  \\
5 &\hst$-1$ &\hst$\kappa_X$ &\hst1.6 &\hst97          &\hst93  &\hst0.51 &\hst0.93 &\hst0.915 &\hst2.665 &\hst1.750 &\hst88.4 &\hst0.338 &\hst0.47 &\hst2.26 &\hst-0.15&\hst15  &\hst1.7  \\
5 &\hst$-2$ &\hst$\kappa_0$ &\hst1.6 &\hst439         &\hst436 &\hst1.29 &\hst0.93 &\hst0.975 &\hst3.565 &\hst2.590 &\hst93.6 &\hst0.453 &\hst1.68 &\hst3.80 &\hst0.99 &\hst240 &\hst1.5  \\
5 &\hst$-2$ &\hst$\kappa_X$ &\hst1.6 &\hst144         &\hst144 &\hst0.65 &\hst0.93 &\hst0.929 &\hst2.260 &\hst1.331 &\hst92.0 &\hst0.362 &\hst1.52 &\hst3.39 &\hst0.62 &\hst94  &\hst3.1  \\
5 &\hst$-2$ &\hst$\kappa_0$ &\hst2.0 &\hst309         &\hst 306&\hst0.99 &\hst0.94 &\hst0.950 &\hst2.006 &\hst1.056 &\hst96.2 &\hst0.452 &\hst1.56 &\hst3.67 &\hst0.76 &\hst170 &\hst2.2  \\
5 &\hst$-2$ &\hst$\kappa_X$ &\hst2.0 &\hst137         &\hst134 &\hst0.59 &\hst0.93 &\hst0.924 &\hst1.657 &\hst0.733 &\hst95.3 &\hst0.375 &\hst1.46 &\hst3.35 &\hst0.46 &\hst85  &\hst5.6  \\
\enddata
\vspace{-0.20cm}
\tablecomments{$M$ is the initial mass in units of M$_\odot$.  The opacity treatment is denoted by $\kappa$, with $\kappa_X$ being models with composition-dependent low-T opacity and $\kappa_0$ those without it.  The MLT mixing length parameter is $\alpha_\text{MLT}=\ell_\text{MLT}/H_\text{p}$ where $\ell_\text{MLT}$ is the mixing length and $H_\text{P}$ is the pressure-scale-height.  The number of thermal pulses is $n_\text{TP}$ and the number with dredge-up is $n_\text{3DU}$.  The AGB lifetime is $t_\text{AGB}$ in Myr. $M_\text{core,f}$, $M_\text{f}$, and $M_\text{env,f}$ are the end-of-computation hydrogen-exhausted core, total, and envelope mass in units of M$_\odot$.  $T_\text{bce,max}$ is the maximum temperature in MK at the base of the envelope during the interpulse phase.  Abundances are given as the average in the wind ejecta (assuming the remaining envelope is ejected without any composition change), where $Y$ is mass fraction of helium (initially $Y=0.245$) and abundances for the $Z=0$ models are expressed as if [Fe/H] $=-4$.  R$_\text{CNO}$ is the ratio of total yield of C+N+O to the initial abundance.  C/O$_\text{f}$ is the final carbon-to-oxygen ratio by number in the envelope.  All models were initially scaled-solar but with [O/Fe] $= +0.4$, except for the 5\,M$_\odot$ [Fe/H] $=-1$ model which was not oxygen-enhanced.}
\end{deluxetable*}

\placetable{table_model_summary}

\begin{figure}
\plotone{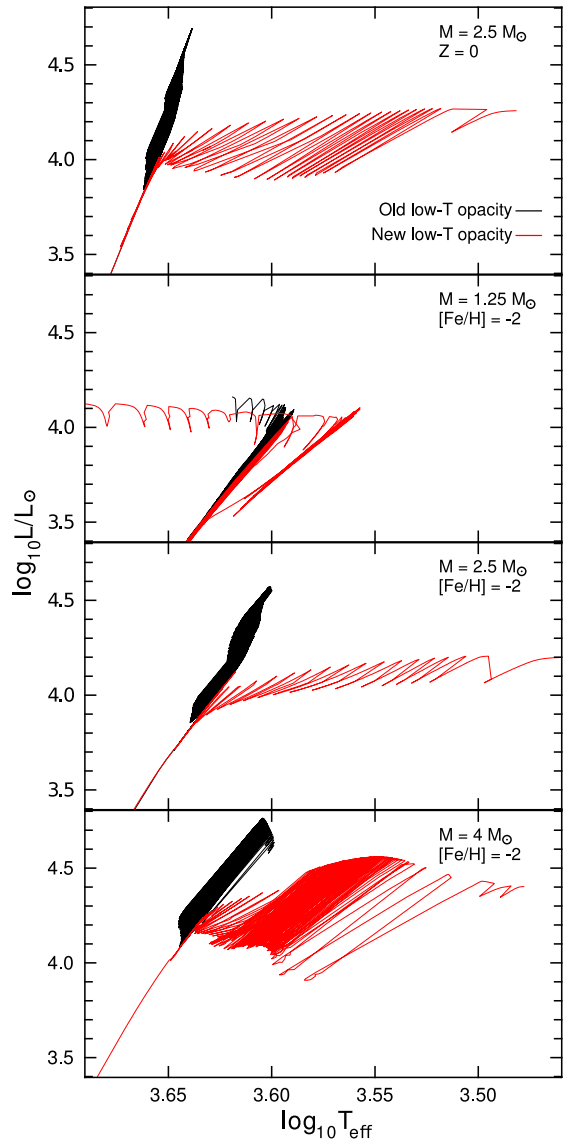}
  \caption{Hertzsprung-Russell (HR) diagrams for representative models.  Models with composition-dependent low-T opacity are in red (grey) and those without it are in black.  The top panel shows 2.5\,M$_\odot$ $Z=0$ models, the remainder are [Fe/H] $=-2$ models with mass 1.25\,M$_\odot$, 2.5\,M$_\odot$, and 4\,M$_\odot$  (from top to bottom).  This color scheme is used throughout. \label{figure_hrds}}
\end{figure}

\placefigure{figure_hrds}

We find that the adoption of composition-dependent low-T opacity alters the evolution of all of our models.  This includes the strongly HBB models (4 and 5\,M$_\odot$), which in their higher-metallicity incarnation have previously been shown to be unaffected \citep{2009MNRAS.399L..54V}.  When the effects become evident, during the third dredge-up, the change is qualitatively what is expected: the additional opacity increases the radius, cools the surface and the base of the envelope, and the mass loss rate increases, shortening the AGB lifetime.  This lifetime reduction means that each model with the new opacity ejects less helium, nitrogen and oxygen.  This reduction in yield is not true for all elements because the conversion of carbon to nitrogen is suppressed in the more weakly HBB models, increasing the carbon yield.  In general, we find that these effects are not metallicity dependent when $\text{[Fe/H]} \leq -2$.  

The results can best be summarized by considering three categories: (i) models that have HBB, (ii) models that no longer have HBB when composition-dependent low-T opacity is used, and (iii) models that do not have HBB.  In this section we also separately analyze our $Z=0$ models.  In Figure \ref{figure_hrds} we show the HR diagram of four representative pairs of these models.  In each example it is apparent that when the new opacity is used the surface becomes cooler instead of hotter during the evolution, and that the maximum AGB luminosity is decreased.

\subsection{Importance for HBB models}
\begin{figure}
\plotone{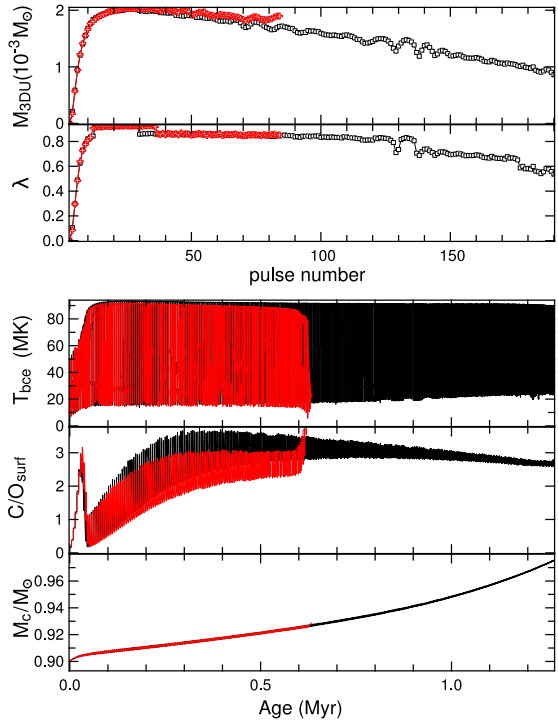}
  \caption{AGB evolution of 5\,M$_\odot$ $\text{[Fe/H]}=-2$ models.  From top to bottom the panels show the mass dredged-up after each thermal pulse, the dredge-up efficiency $\lambda$, the temperature at the base of the convective envelope, surface C/O number ratio, and hydrogen-exhausted core mass.  The age has been set to zero at the beginning of the TP-AGB phase.  The colors are the same as in Figure \ref{figure_hrds}.\label{figure_m5z0001_3DU}}
\end{figure}

\begin{figure}
\plotone{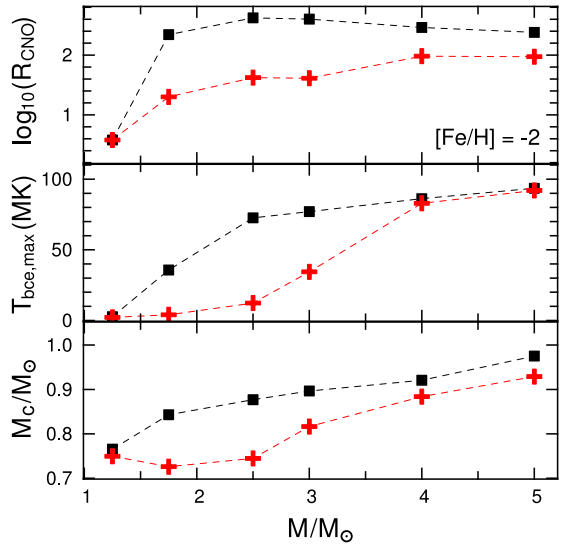}
\caption{Comparison of the [Fe/H] $=-2$ models.  Upper panel: Ratio of the total yield of C+N+O to the initial abundance (R$_\text{CNO}$).  It is assumed that the remaining envelope is ejected with the same composition as at the end of computation.  Middle panel: Maximum temperature at the base of the convective envelope during the interpulse period for the same models. Lower panel: End of computation hydrogen-exhausted core mass ($M_\text{c}$).  The colors are the same as in Figure \ref{figure_hrds}.\label{figure_summary_mass} }
\end{figure}

In HBB models ($M > 3$\,M$_\odot$) with $\text{[Fe/H]} \leq -2$, composition-dependent low-T opacity causes the AGB evolution to be truncated because of faster mass loss.  The accelerated mass loss reduces the number of thermal pulses by a factor of three and the AGB lifetime by a factor of two (Figure \ref{figure_m5z0001_3DU}).  The maximum temperature at the base of the convective envelope ($T_\text{bce,max}$) is slightly reduced, by about $1 - 3$\, MK, from roughly $86 - 96$\,MK (Table \ref{table_model_summary}).  The mean mass fraction of helium $Y$ in the ejecta is reduced by around 0.1, to approximately $Y=0.35$, which has consequences for the GC abundance pattern problem (discussed in Section \ref{subsec:GC}).  The total enrichment in C+N+O is reduced by a factor of two to three (Figure \ref{figure_summary_mass}).  The main component in this is the nitrogen yield, which is more than halved in the  $\text{[Fe/H]}\leq -2$ models (Table \ref{table_model_summary}).  In these models, dredge-up efficiency and core growth rate are unaffected by the change in surface conditions (Figure \ref{figure_m5z0001_3DU}).  In our 5\,M$_\odot$ $\text{[Fe/H]}=-1$ model the evolution difference is less extreme but still clear.  This shows that compared to \citet{2010MNRAS.408.2476V}, HBB models using the physics in our code are sensitive to the low-T opacity treatment up to a higher metallicity.  In fact, the inclusion of the new opacity always makes a significant difference to the AGB evolution, so it is necessary for every one of the HBB models examined.

The evolution of our 4\,M$_\odot$ models with $\text{[Fe/H]} = -2$ and $\text{[Fe/H]}=-3$ is very similar.  It is difficult to determine if the effect of the new opacity is metallicity-dependent because the models with the old opacity still had significant envelope mass at the end of computation.  The two models with the new opacity are possibly more alike than those without: the [Fe/H] $=-3$ model has 103 thermal pulses and an AGB lifetime of 0.77 Myr compared to 85 thermal pulses and 0.65 Myr for the [Fe/H] $=-2$ model, with very similar chemical yields (Table \ref{table_model_summary}).  In contrast, the envelope of the lower-metallicity model with the old opacity was more helium-rich than the higher metallicity model ($Y = 0.462$ compared to $Y = 0.442$) at the end of computation.  The former model also had more envelope still remaining (2.9\,M$_\odot$ compared to 1.2\,M$_\odot$), and would therefore become even more enriched in helium.

\begin{figure}
\plotone{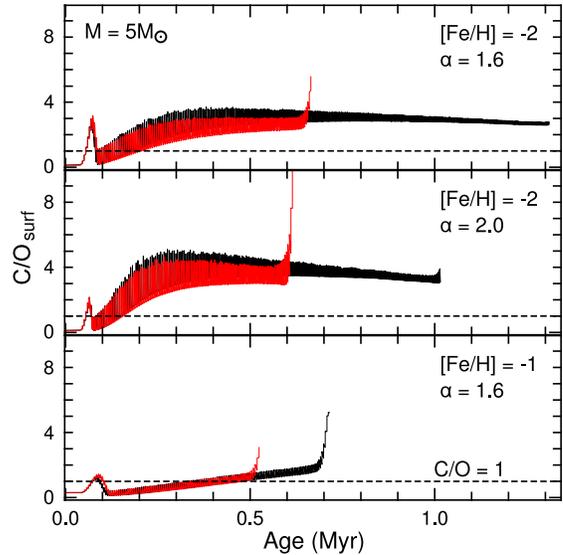}
  \caption{Surface C/O ratio (by number) for each of the 5\,M$_\odot$ models.  The models in the top two panels have metallicity $\text{[Fe/H]}=-2$, while the model in the bottom panel has $\text{[Fe/H]}=-1$.  The model in the middle panel has MLT mixing length parameter $\alpha_\text{MLT} = 2$ instead of 1.6).  The colors are the same as in Figure \ref{figure_hrds}.\label{figure_m5_co}}
\end{figure}

Interestingly, every one of our HBB models attains $\text{C/O} > 1$ during the evolution (Figure \ref{figure_m5_co}).  This initially occurs quickly after carbon is dredged-up, while the temperature at the base of the convective envelope ($T_\text{bce}$) is still increasing.  By the time HBB becomes established the additional opacity has already had an effect -- the radius of the 4\,M$_\odot$ $\text{[Fe/H]} = -2$ model with the old opacity is 60 R$_\odot$ larger than its counterpart with the old opacity (which has radius around 350\,R$_\odot$).  The envelope carbon abundance then rapidly falls (C/O reduces to below 1 in the $\text{[Fe/H]} \ge -2$ models) before C/O slowly rises again and remains above unity.  By comparison, the 3.5\,M$_\odot$ $\text{[Fe/H]}\simeq -1.5$ models from \citet{2009MNRAS.399L..54V} do not become carbon-rich.  This difference may be attributed to a number of causes. Firstly, we use a different convection formalism, MLT instead of the full spectrum of turbulence, and have more efficient carbon dredge-up. Secondly, our models have a lower initial oxygen abundance (the $\text{[Fe/H]}= -1$ model is not oxygen-enhanced and the rest are lower metallicity).  Lastly, our mass loss rate is slower.  The most direct comparison we can make to \citet{2010MNRAS.408.2476V} is between their 5\,M$_\odot$ $\text{[Fe/H]}\simeq -1.5$ and our 5\,M$_\odot$ $\text{[Fe/H]}=-1$ model.  The core masses are comparable (within 0.04\,M$_\odot$ at most), with ours slightly smaller, giving a marginally lower temperature at the base of the convective envelope (88\,MK compared to roughly 105\,MK).  The main difference is our larger number of thermal pulses (97 compared to their 33 and 54 when composition-dependent low-T opacity is used).  This longer AGB lifetime, with more thermal pulses, allows for more 3DU in our models and eventually $\text{C/O} > 1$.  The main reason behind this is our slower mass loss rate compared to the \citet{1995A&A...297..727B} and \citet{2006NuPhA.777..311S} rates used by \citet{2010MNRAS.408.2476V}.  This allows for additional envelope enrichment and consequently an opacity increase when the new opacity is used.

\placefigure{figure_m5z0001_3DU}

\placefigure{figure_summary_mass} 

\placefigure{figure_m5_co}

\subsection{Borderline HBB models}
\begin{figure}
\plotone{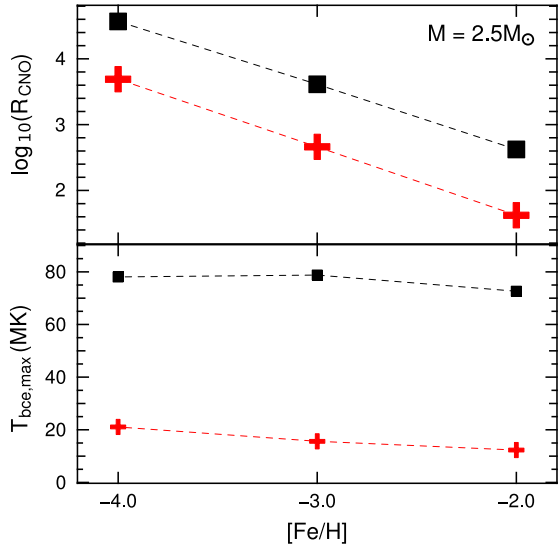}
\caption{Upper panel: Ratio of the total yield of C+N+O to the initial abundance (R$_\text{CNO}$) for each of the 2.5\,M$_\odot$ models.  It is assumed that the remaining envelope is ejected with the same composition as at the end of computation.  Lower panel: Maximum temperature at the base of the convective envelope during the interpulse period.  The colors are the same as in Figure \ref{figure_hrds}. \label{figure_summary_FeonH}}
\end{figure}

\begin{figure}
\plotone{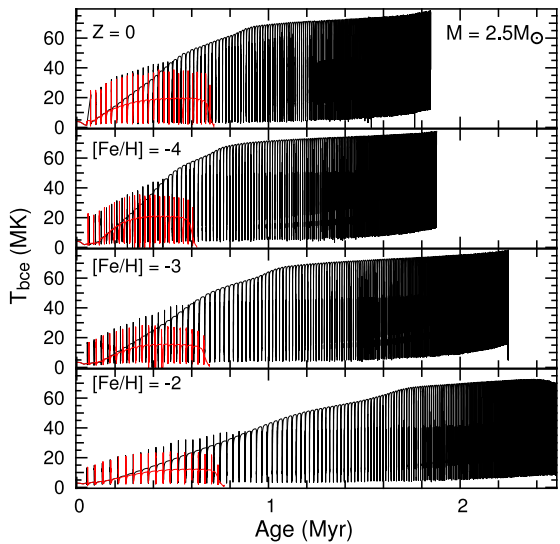}
  \caption{Temperature at the base of the convective envelope for each of the 2.5\,M$_\odot$ models.  They have metallicity $Z=0$, $\text{[Fe/H]}=-4$, $\text{[Fe/H]}=-3$, and $\text{[Fe/H]}=-2$, from top to bottom.  The colors are the same as in Figure \ref{figure_hrds}.\label{figure_M25_Tbce}}
\end{figure}

In our lower-mass models ($1.75 \lesssim \text{M}/\text{M}_\odot \lesssim 3$) hot bottom burning is ``quenched,'' as it is for the higher-metallicity models in \citet{2010MNRAS.408.2476V}.  We do not see a strong dependence of $T_\text{bce}$ on metallicity (Figure \ref{figure_summary_FeonH}), nor do we find a threshold metallicity below which HBB is no longer suppressed by the use of the new opacity (Figure \ref{figure_M25_Tbce}).  On the contrary, in our metal-poor $\text{[Fe/H]}=-2$ models we find HBB suppression between about 1.75\,M$_\odot$ and 3\,M$_\odot$.  At this metallicity the 1.75\,M$_\odot$ composition-independent low-T opacity and the 3\,M$_\odot$ composition-dependent low-T opacity models both reach the same maximum $T_\text{bce}$ of about 35\,MK (Table \ref{table_model_summary}).  We can therefore accurately quantify the increase in the HBB threshold to be $\Delta M_\text{HBB} =  1.25$\,M$_\odot$.  This appears to be a wider mass range compared to \citet{2010MNRAS.408.2476V}, both for the models with the \citet{2006NuPhA.777..311S} mass loss rate and (certainly) for the models with the \citet{1995A&A...297..727B} rate.  It therefore appears that our slower mass loss rate (particularly earlier on the AGB when the pulsation period is shorter), which terminates the AGB earlier, can account for some of the difference between these models.  

\begin{figure}
\plotone{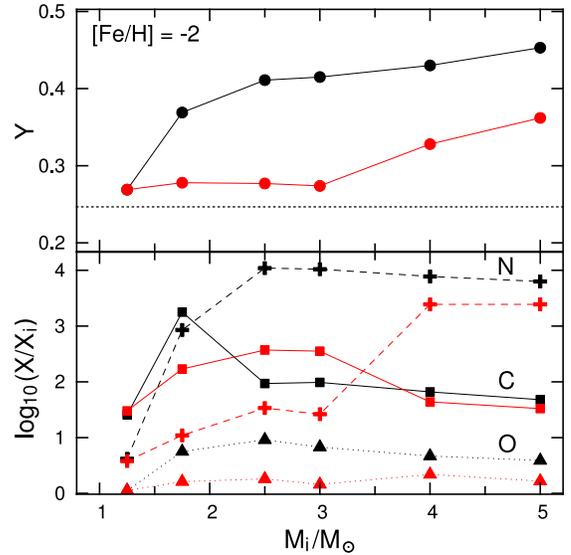}
  \caption{Integrated yields for each [Fe/H] $=-2$ model.  Upper panel: Average mass fraction of helium in ejecta.  The dashed line is initial helium abundance $Y=0.245$.  Lower panel: Carbon (square), nitrogen (cross) and oxygen (triangle) yields expressed as a log fraction of initial abundance.  It is assumed that any remaining envelope is ejected with the end-of-computation abundance.  The colors are the same as in Figure \ref{figure_hrds}.\label{figure_FeonH-2_yields}}
\end{figure}

The evolution is so divergent in this mass range because of a feedback process.  The additional opacity stops HBB which allows the carbon abundance to increase, causing a further increase in opacity, then expansion, cooling and mass loss.  The shortened lifetime due to the faster mass loss then prevents the H-exhausted core from growing (and the consequential increase in $T_\text{bce}$).  Compared to the models without the new opacity, the AGB lifetime is reduced by a factor of three to four, causing the largest reduction in final core mass (Figure \ref{figure_summary_mass}), which is greater than 0.1\,M$_\odot$ in all of these models. The number of thermal pulses and the total yield of C+N+O are reduced by a factor of 10 compared to equivalent composition-independent opacity models.  The carbon yield is increased by up to a factor of 10 whereas the nitrogen and oxygen yields are reduced by around 99.7\% and 80\% respectively.  This major change in composition over such a broad mass range may have important implications for the predicted CEMP and NEMP frequency (discussed in section \ref{subsec:CEMP}).  These models also have the largest reduction in helium production, with a mass fraction difference of up to 0.16 (Figure \ref{figure_FeonH-2_yields}).

\placefigure{figure_summary_FeonH}

\placefigure{figure_M25_Tbce}

\placefigure{figure_FeonH-2_yields}

\subsection{Low-mass models}
\begin{figure}
\plotone{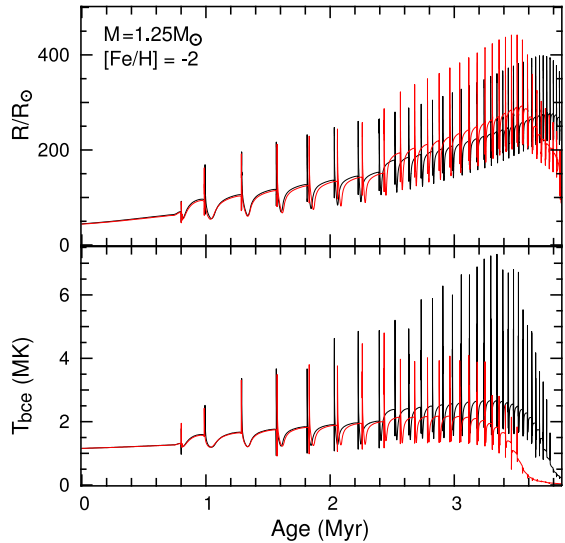}
  \caption{Stellar radius (upper panel) and temperature at the base of the convective envelope (lower panel) for the 1.25\,M$_\odot$ $\text{[Fe/H]}=-2$ models.  The colors are the same as in Figure \ref{figure_hrds}.\label{figure_M125_Teff_R}}
\end{figure}

In the mass range below which HBB can occur, the results are critically dependent on the extent of third dredge-up.  Both of our 1.25\,M$_\odot$ models have only one small third dredge-up episode, achieving $\text{C/O} \approx 3$ and almost identical final yields. The relatively modest change in surface composition that results, and the already fast mass loss, mean that the structure and lifetime of the new and old opacity cases are more alike than for the other models studied.  Despite this, the change in structure after the dredge-up is still obvious: in the model with the new opacity the radius increases substantially and there is a slower increase in $T_\text{bce}$ (Figure \ref{figure_M125_Teff_R}).  Even though this model initially had $\text{[O/Fe]}=+0.4$ and only experiences a single, low-efficiency third dredge-up episode ($\lambda = 0.17$), it easily reaches $\text{C/O} > 1$.  This suggests that \textit{any} low-mass, low-metallicity model with 3DU requires composition-dependent low-T opacity.

\begin{figure}
\plotone{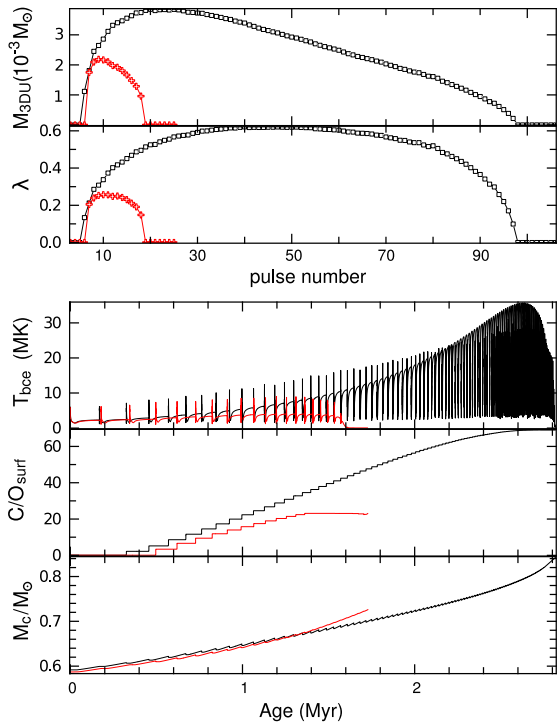}
  \caption{AGB evolution of 1.75\,M$_\odot$, [Fe/H] $=-2$ models.  From top to bottom the panels are the mass dredged-up each thermal pulse, the dredge-up efficiency $\lambda$, the temperature at the base of the convective envelope, surface C/O number ratio, and hydrogen-exhausted core mass.  The colors are the same as in Figure \ref{figure_hrds}.\label{figure_m175z0001_3DU}}
\end{figure}

It is evident that we cannot characterize low-mass composition-dependent low-T opacity models ($\text{[Fe/H]}=-2$) with substantial 3DU (e.g. the $M=1.75$\,M$_\odot$ model) as truncated copies of models without the new opacity.  Differences in the structure are evident from the beginning of the TP-AGB.  The additional opacity leads to increased mass loss and then a lower $T_\text{bce}$ and 3DU efficiency (Figure \ref{figure_m175z0001_3DU}).  The reduced dredge-up and shorter lifetime give the same reduction in total C+N+O yield as for the models where the new opacity causes HBB to be averted.  Qualitatively, these effects are similar to those for higher-metallicity models.  The opacity treatment causes a more pronounced divergence in evolution in our 1.75\,M$_\odot$ than it does for the 2.5\,M$_\odot$ $\text{[Fe/H]}\simeq -1.5$ models in \citet{2009MNRAS.399L..54V} (which is an apt comparison because they are both just below the mass cut-off for HBB in the respective codes).   We find a maximum $T_\text{bce}$ of 36\,MK and 4\,MK compared to 31\,MK and 18\,MK in \citet{2009MNRAS.399L..54V}.  Similarly, the differences in chemical yields, lifetime and number of thermal pulses  in our models are larger.  Although the use of the \citet{1995A&A...297..727B} mass loss certainly contributes to the contrast between the two sets of models, it also highlights that the lower the metallicity, the more acute is the need for the new opacity.

The only comparable study for a metal-poor case that we could find in the literature is the 2\,M$_\odot$ $\text{[Fe/H]}= -2.17$ model in \citet{2007ApJ...667..489C}.  In their model, accounting for C and N enhancements reduced the number of thermal pulses by a factor of four.  This compares to a factor of three for our 1.75\,M$_\odot$ $\text{[Fe/H]}=-2$ model.  The halting of growth of the third dredge-up is also similar to our model.  The main difference between the models is that ours have approximately double the number of thermal pulses (106 and 25 compared to 51 and 15).  Their use of the \citet{2006NuPhA.777..311S} mass loss rate contributes to this, because it is higher from the beginning of the AGB, when the pulsation period $P$ is shorter ($\log{\left[P \text{(days)}\right]}<2.5$).  In fact, our model with the new opacity only reaches this pulsation period after 48 thermal pulses and 80\% of its TP-AGB lifetime.

\placefigure{figure_M125_Teff_R}

\placefigure{figure_m175z0001_3DU}

\subsection{Lessons from zero-metallicity models} \label{subsec:zero}
Zero-metallicity models are interesting in this study because they display the most extreme change in composition during their evolution.  We examined a 2.5\,M$_\odot$ $Z=0$ model in order to determine if there is indeed a metallicity limit below which the HBB ``quenching'' effect of the new opacity is no longer observed.  This model presented an additional complication because it is the only one in our grid in which the intershell convection zone is able to penetrate the H-shell at the beginning of the TP-AGB, leading to an event that is variously referred as an H-flash, dual shell flash, or proton ingestion episode \citep{1996ApJ...459..298C,2008A&A...490..769C,2009PASA...26..145I}.  In 1D models this can radically alter the surface composition.  Studies by \citet{2009PASA...26..145I} and \citet{2010MNRAS.405..177S}, however, show that the ensuing surface enrichment decreases with increasing stellar mass.  In the $Z=0$ models from the latter, the total C+N+O fraction in the 2\,M$_\odot$ model is a factor of 20 less than the 1.5\,M$_\odot$ model.  While in models with $M\geq2$\,M$_\odot$ there is some mixing to the surface from this event it is relatively minor and the envelope composition is very quickly dominated by the subsequent third dredge-up events and HBB (see e.g. the 3\,M$_\odot$ $\text{[Fe/H]}=-5.45$ model and summary schematic in \citealt{2008A&A...490..769C}, their Figures 2 and 4).   Overall, the dual shell flash is difficult to model and the results are highly uncertain in 1D
codes. Importantly, it is unlikely to be relevant to our opacity study because we expect little effect on the later evolution of a 2.5\,M$_\odot$ model.  For these reasons we began the comparison with a single TP-AGB 2.5\,M$_\odot$ $Z=0$ model after the first thermal pulse (i.e. we did not evolve \emph{both} the new and old opacity models from the pre-MS), in which we specifically prohibited the intershell convection zone from expanding into the H-shell.

With the old opacity the final AGB core mass is much larger ($\Delta M_\text{c,f}= 0.13$\,M$_\odot$, which is consistent with the higher metallicity 2.5\,M$_\odot$ models  in Table \ref{table_model_summary}).  Like the other models with the new opacity, the evolution is hastened via positive feedback: the initial cooling of the envelope prevents HBB, so dredged-up carbon is not converted to nitrogen  (which contributes less to the opacity).  With the new opacity, the mass limit for zero-metallicity HBB models appears to be well over 2.5\,M$_\odot$, compared to 2\,M$_\odot$ in \citet{2002ApJ...570..329S}.  This contrasts strongly with \citet{2008A&A...490..769C} where every one of the $Z=0$ models ($0.8 \leq M/\text{M}_\odot \leq 3$) produced more nitrogen than carbon, and the 2 and 3\,M$_\odot$ models both have $\text{N/C} > 20$.  This ratio is reversed for our 2.5\,M$_\odot$ model with the new opacity.  Since the effect of the new opacity is evident in the most metal-poor case possible, these findings demonstrate that there is no lower metallicity limit below which the new opacity is not essential if carbon is dredged-up.  An examination of the {\sc aesopus} data reveals that while the opacity is less dependent on composition at this model's relatively high surface temperature ($\log{T} \sim 3.7$), it is still much more sensitive to an increase in carbon than it is to nitrogen.

\section{Discussion}
\subsection{Key findings}
\begin{figure}
\plotone{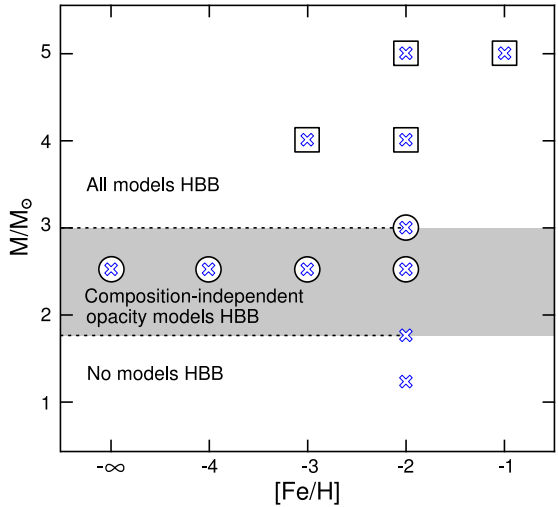}
  \caption{Results from the grid of stellar models.  Each model pair (with and without composition-dependent opacity) is marked by a cross.  Circles indicate that HBB only occurs in the model without composition-dependent opacity.  Squares indicate that HBB occurs in both models.  The shaded region is the approximate mass range that only has HBB when composition-independent opacity is used.  The dashed lines indicate the metallicity range studied $\text{[Fe/H]} \leq -2$ and where the evolution depends little on the initial metallicity.\label{figure_HBB_summary}}
\end{figure}

\begin{figure}
\plotone{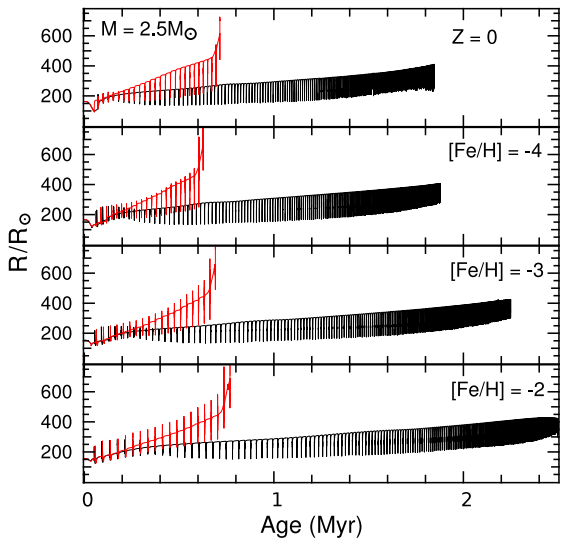}
  \caption{Stellar radius for each of the 2.5\,M$_\odot$ models.  They have metallicity $Z=0$, [Fe/H] $=-4$, [Fe/H] $=-3$, and [Fe/H] $=-2$ from top to bottom.  The colors are the same as in Figure \ref{figure_hrds}.\label{figure_M25_R}}
\end{figure}

Our models do not support the suggestion by \citet{2007A&A...467.1139M} that there is a metallicity limit below which composition-dependent low-T opacity may be neglected (Figure \ref{figure_HBB_summary}).  In fact, as metallicity decreases there is remarkable uniformity in the differences in radius (Figure \ref{figure_M25_R}), $T_\text{bce}$ (Figure \ref{figure_M25_Tbce}), and the total yield of C+N+O relative to the initial abundance (R$_\text{CNO}$; Figure \ref{figure_summary_FeonH}).  Moreover, the absolute mass fractions of C, N, and O in the stellar wind from models differing only in initial metallicity are almost identical.  We also find the new opacity to be crucial for models massive enough to have HBB.

\placefigure{figure_HBB_summary}

\placefigure{figure_M25_R}

\subsection{Implications for CEMP and NEMP stars}
\label{subsec:CEMP}
\begin{figure}
\plotone{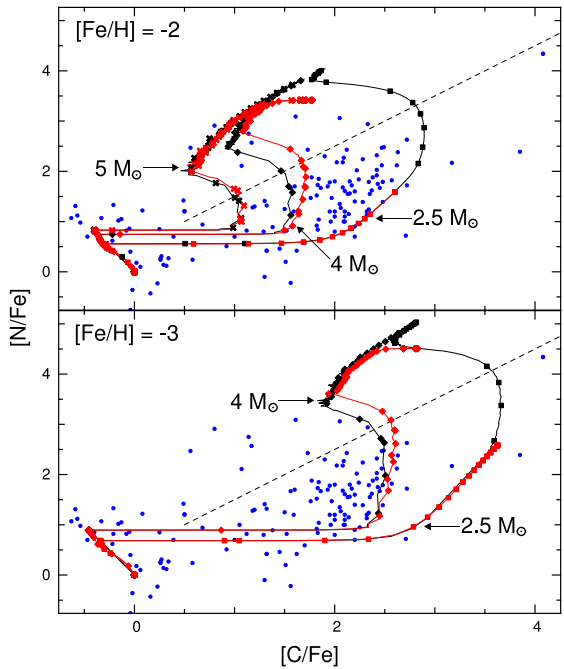}
  \caption{Upper panel: Surface abundance of nitrogen and carbon for 2.5\,M$_\odot$, 4\,M$_\odot$, and 5\,M$_\odot$ [Fe/H] $=-2$ models (in squares, diamonds, and crosses respectively) plotted during the evolution at intervals of 10\,R$_\odot$ (instead of time).  Models with composition-dependent low-T opacity are in red (grey) and those without it are in black.  Lower panel: Same as upper panel except for 2.5\,M$_\odot$ and 4\,M$_\odot$ [Fe/H] $=-2$ models (in squares and diamonds respectively).  The blue circles are observations listed in \citet{2010A&A...509A..93M} which have $\text{[Fe/H]} \leq -0.99$ and a mean of $\text{[Fe/H]}=-2.74$.  The dashed line is [N/C]=0.5 and [C/Fe] $>0.5$, which separates CEMP and NEMP stars according to the definition in \citet{2012A&A...547A..76P}. \label{figure_NCEMP} }
\end{figure}

Our results go some way to explaining the paucity of NEMP stars.  In the entire mass and metallicity range explored, models with composition-dependent low-T opacity produce less nitrogen than the models without it.  Lifetimes are always shortened, leading to fewer dredge-up events and shorter HBB time.  The higher-mass, low-metallicity models that produce the most nitrogen have their AGB lifetime reduced by around two thirds (Table \ref{table_model_summary}).  Moreover, at lower masses the cooling effect in the interior is sufficient to prevent altogether the conversion of the carbon in the envelope to nitrogen before it is shed in the stellar wind (Figure \ref{figure_NCEMP}).  This causes an increase in the mass threshold for HBB of $\Delta M_\text{HBB} =  1.25$\,M$_\odot$ in our [Fe/H] $=-2$ models.  With an IMF that favors low-mass stars (\citealt{2012A&A...547A..76P} exclude the possibility of a top-heavy IMF for $-2.8 \leq \text{[Fe/H]} \leq -1.8$) we predict that such a change would give a significant reduction in the expected NEMP/CEMP ratio in a population synthesis model.  There are more consequences to consider here too.  Not only is the composition of the wind different, but also the evolution of radius with time, which would obviously affect Roche-lobe overflow mass transfer and common envelope evolution in binary systems.  In general, the increase in maximum radius, which can result from a single 3DU episode (Figure \ref{figure_M125_Teff_R}), would increase the likelihood of binary interaction.

The existence of NEMP stars only below $\text{[Fe/H]} =-2.8$ is still puzzling.  Although it is known that $T_\text{bce}$ increases with decreasing metallicity for a given MLT mixing length \citep{1991ApJ...366..529S}, and strengthens HBB, the difference in particular between our $\text{[Fe/H]} = -2$ and $\text{[Fe/H]} = -3$ models (Figure \ref{figure_M25_Tbce}) cannot explain the observed [Fe/H] dependence of NEMP star existence \citep{2007ApJ...658.1203J}.   The new opacity appears not to help the situation: the magnitude of the change in $T_\text{bce}$ with a change in metallicity is independent of the low-T opacity treatment.   Comparing our 2.5\,M$_\odot$ $\text{[Fe/H]} = -2$ and $Z=0$ models illustrates this point.  In the new opacity case the maximum $T_\text{bce}$ increases from 12.3 to 19.7 MK with the drop in metallicity, while the composition-independent case increases from 72.6 to 78.5 MK (Table \ref{table_model_summary}).  Our new and old opacity 4 and 5\,M$_\odot$ models show almost the same drop in $T_\text{bce}$ as a result of a reduction in metallicity $\Delta\text{[Fe/H]}=-1$ (to within 0.2 and 1.2 MK respectively).  

Our models cannot explain the metallicity dependence of the existence of NEMP stars through binary accretion.  Assuming there is indeed a need to further reduce the HBB limit only at very low metallicity to reproduce the observed frequency of NEMP stars (i.e. the nitrogen-enhancement is not the result of another process), a metallicity or composition-dependent mass loss rate may offer an explanation, perhaps relating to dust formation \citep{2012A&A...547A..76P}.  The \citet{1993ApJ...413..641V} rate we use is only dependent on stellar structure and therefore becomes relatively insensitive to [Fe/H] for extremely metal-poor models.  Alternatively, an IMF more heavily weighted to intermediate-mass stars at very low metallicity would help \citep{2013MNRAS.432L..46S}.

Our low-mass models computed with the new opacity treatment have a much shorter TP-AGB phase and reduced oxygen yield.  In Figure 4 in \citet{2011AJ....141..102K} the observed [O/Fe] in 19 CEMP stars is compared to the average value for EMP stars from \citet{2005A&A...430..655S}.  Although the error bars are quite large ($\gtrsim 0.5$ dex), the two groups are comparable.  This suggests the reduced oxygen enhancement in models with the new opacity is reasonable.   In that figure there is also an [O/Fe]-[Fe/H] relation which we replicate with our 2.5\,M$_\odot$ models using \emph{either} opacity treatment.  

\begin{figure}
\plotone{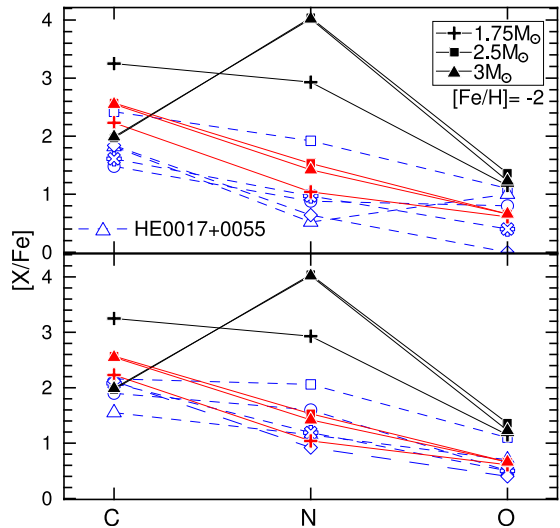}
  \caption{Integrated chemical yields of C, N, and O from 1.75, 2.5, and 3\,M$_\odot$ [Fe/H] $=-2$ models (solid symbols) compared to ten observed CEMP stars from Figure 7 of \citet{2011AJ....141..102K} (in blue, open symbols and dashed lines).  Models with composition-dependent low-T opacity are in red (grey) and those without it are in black.  The observations are split between two panels, consistent with the original figure.  HE0017+0055 is the only star that the models with the new opacity cannot match (see text for details).\label{figure_Kennedy}}
\end{figure}

In Figure \ref{figure_Kennedy} we compare the C, N, and O yields from our 1.75, 2.5, and 3\,M$_\odot$  $\text{[Fe/H]}=-2$ models to the ten stars in \citet{2011AJ....141..102K} for which these abundances were determined.  Each of these stars has $\text{[C/N]} > 0$.  This immediately rules out the $M$/M$_\odot > 1.75$ models with the old opacity from matching a donating companion, because they produce far too much N from HBB.  With the exception of the very O-rich star HE0017+0055, all stars can be matched reasonably (to within about 0.5 dex) by the three new opacity models.   The abundance patterns in the new opacity models are very similar to three of the observed stars.  A further six observed stars are closer to having [C/Fe] = [N/Fe].  These six observations are consistent with our models if we consider the scenario in \citet{2007A&A...464L..57S} where thermohaline mixing of accreted material is followed by the first dredge-up which reduces the initially high surface [C/Fe].  In \citet{2007A&A...464L..57S} the model of a 0.74\,M$_\odot$ secondary accreting 0.09\,M$_\odot$ eventually reaches $\text{[C/Fe]} - \text{[N/Fe]} < 0.5$, irrespective of initial N accretion.  Surprisingly, nine out of the 10 observed abundance patterns are then better explained by the new opacity models (Figure \ref{figure_Kennedy}; the very O-rich star HE0017+0055 cannot be matched by any of our models).  The new opacity models are also a good fit for the observations plotted in the [C/Fe]-[O/Fe] plane in Figure 8 in \citet{2011AJ....141..102K} after dilution is taken into account.

\citet{2009A&A...508.1359I} suggest that more efficient dredge-up for $M < 1.25$\,M$_\odot$ in binary population synthesis models is needed to fit the higher observed CEMP/EMP ratio.  While the updated low-T opacity certainly does not solve this problem (it is not even relevant unless there is dredge-up), it would be interesting to study the effect of low-T opacity in low-mass models with more efficient dredge-up.

\subsection{AGB stars as globular cluster polluters}
\label{subsec:GC}
The 5\,M$_\odot$ [Fe/H] $=-2$ models with composition-dependent low-T opacity yield only half the oxygen and total C+N+O.  Although this change is an improvement in the right direction, it is not nearly enough for these models to be consistent with the intermediate-mass AGB globular cluster self-pollution scenario.  Evidence points towards a spread in the helium abundance within globular clusters, and that the extent of the spread differs immensely from cluster to cluster.   Fitting stellar evolution isochrones to main sequence, subgiant and red giant branch HST photometry suggests there is a small helium spread $\Delta Y \sim 0.03$ in NGC 6752 \citep{2013ApJ...767..120M}.  In NGC 2808, by contrast, isochrone fits of the horizontal branch \citep{2005ApJ...621L..57L} and main sequence \citep{2005ApJ...631..868D,2007ApJ...661L..53P} along with spectroscopic determination \citep{2011A&A...531A..35P} show a much larger helium spread $\Delta Y \sim 0.20$.   All of our 5\,M$_\odot$ models become enriched in helium (Figure \ref{figure_FeonH-2_yields}), regardless of opacity treatment, because much of the increase results from the second dredge-up, which occurs before the new opacity treatment has any effect.  The stellar wind in the most helium-rich model with the new opacity, however, has $Y=0.375$ ($\Delta Y = 0.13$) which is insufficient to account for the helium-rich population in NGC 2808.

Even when the new opacity is used for the $\text{[Fe/H]}=-2$ models we still do not find any oxygen depletion.  The mean ejected [O/Fe] is 0.62 and 0.46 in the standard and higher MLT mixing length cases respectively (Table \ref{table_model_summary}).  This compares to approximately scaled-solar oxygen in Na-rich stars observed in 5 clusters of similar metallicity $-2.5<\text{[Fe/H]}<-1.8$ \citep{2012A&A...539A..19G}.   This naturally points towards further increasing the mixing length to generate models that fit the [O/Fe] constraint.  A problem is that in our mixing length test (last two models in Table \ref{table_model_summary}) R$_\text{CNO}$ is barely affected (reducing from 94 to 85 when increasing $\alpha_\text{MLT}$ from 1.6 to 2.0 for the models with the new opacity), and is still considerably higher than the observational constraints (summarized in the Introduction).  Faster mass loss would then be required to limit the total 3DU mass.  We note that \citet{2013MNRAS.433..366D} used a recent version of our code to reveal some additional problems with the above approach when also performing detailed nucleosynthesis calculations to replicate the abundance patterns in NGC 6121 (M4).  Our higher metallicity ($\text{[Fe/H]}=-1$) 5\,M$_\odot$ models did deplete oxygen, producing ejecta with $\text{[O/Fe]} =-0.15$ (irrespective of the opacity treatment) but not enough to explain the oxygen spread observed by \citet{2012A&A...539A..19G}. R$_\text{CNO}$ was also too high, being 15 in the model with the new opacity and 21 in the model without it.

\subsection{Uncertainties}
The general trends we have established are dependent on the mass loss prescription, which is poorly constrained for low-metallicity AGB stars.  The difference in lifetimes would be smaller, for example, if a higher mass loss rate were adopted.  When the new opacity is used in our models the mass loss rate is very heavily dependent on envelope enrichment, both compared to models without it and models with other mass loss prescriptions.  Mass loss formulae principally depend on stellar mass, luminosity and radius to different degrees.  The quantity most directly affected by the new opacity is radius.  Therefore, the changes are most obvious when using the mass loss formula more sensitive to radius.  According to \citet{1975MSRSL...8..369R} and \citet{1995A&A...297..727B}, mass loss is proportional to radius $R$, compared to \citet{1990A&A...231..134N} where it is proportional to $R^{0.81}$.  This contrasts with \citet{1993ApJ...413..641V} in which the mass loss rate increases exponentially with $R^{1.94}$ (prior to the superwind phase).  This formula uses the pulsation period dependence on mass and radius originally computed for low-mass models  $0.6 \lesssim M/\text{M}_\odot \lesssim 1.5$ \citep{Wood1989}, and it also does not directly account for the change in stellar structure caused by the additional low-T opacity.  Even so, we believe that using the \citet{1993ApJ...413..641V} rate is a reasonable choice because after the third dredge-up the Z content of the envelope and the stellar structure is comparable to the more metal-rich stars from which period-mass loss relation was empirically derived.  

We investigated the effect of the MLT mixing length, which could be important because it affects the two factors that control the conditions at the surface, the temperature and chemistry.  This includes its effect on the conditions at the interior boundary of the convective envelope that control dredge-up and HBB, and thus alter the surface composition.  In our comparison between 5\,M$_\odot$ models with mixing length parameter $\alpha_\text{MLT}=\ell_\text{MLT}/H_\text{p}$ of 1.6 and 2.0 (where $\ell_{MLT}$ is the mixing length and $H_\text{P}$ is the pressure-scale-height) we find only minimal difference, apart from the expected shift to slightly higher $T_\text{eff}$ with increasing $\alpha_\text{MLT}$ (Table \ref{table_model_summary}).  It therefore appears that our analysis is insensitive to small changes in mixing length, at least for HBB models.  

Dredge-up efficiency $\lambda$, which depends on the treatment of convection and overshoot, is another important consideration for how models are affected by the low-T opacity.  This is because it leads to important changes in the structure and composition: the core growth rate is higher in models with little dredge-up, leading to increased luminosity according to the well-known relation for non-HBB models \citep{1970AcA....20...47P}, and therefore faster mass loss, while envelope enrichment will be slower.   These two factors combine so that the evolution of models with a higher $\lambda$ will be more affected by the new opacity than otherwise similar models with a lower $\lambda$.  A comparison in Figure 4 of \citet{2013MNRAS.434..488M} between five full stellar structure codes \citep{2011ApJS..197...17C,2009A&A...508.1343W,2004MNRAS.352..984S,2002PASA...19..515K,2000A&A...360..952H} with the same initial model (3\,M$_\odot$ $Z=0.02$) shows how much $\lambda$ and the minimum core mass for 3DU can vary from code to code.  The dredge-up efficiency is important in HBB models for an additional reason: C/O depends on the competition between carbon dredge-up and its destruction.  Our HBB models tend to have a high $\lambda$ (peaking between 0.91 and 0.99 in $M > 2.5$\,M$_\odot$ models with the old opacity) which partly explains why we find the opacity to be so important where others have not \citep[e.g.][]{2010MNRAS.408.2476V}.  While the third dredge-up is crucial to determining the impact of the new opacity, the increase in computed opacity itself has little or no effect on $\lambda$ (Table \ref{table_model_summary}) except when the mass loss rate increases, terminating the AGB evolution earlier (Figure \ref{figure_m5_co}).  \citet{2010MNRAS.408.2476V} attribute the three-fold difference in the number of thermal pulses between their 2\,M$_\odot$ $Z=0.001$ model and an equivalent in \citet{2009A&A...508.1343W} to the latter's deeper 3DU.  This causes the envelope to become C-rich earlier and speeds up mass loss.  Low-metallicity models become C-rich very easily (our 1.25\,M$_\odot$ and 1.75\,M$_\odot$ $\text{[Fe/H]}=-2$ both had $\text{C/O}>1$ after the first 3DU episode) so this factor is less important in our low-mass models.

\section{Conclusions}
We find that the inclusion of composition-dependent low-T opacity influences the evolution of all of our models.  Although the resulting structural effects for metal-poor models are broadly similar to those reported elsewhere for higher-metallicity models \citep[e.g.][]{2002A&A...387..507M,2009MNRAS.399L..54V}, we find them to be applicable over a broader stellar mass range, at least $1.25 \leq M/\text{M}_\odot \leq 5$.  In metal-poor models the third dredge-up can more easily increase the surface metal abundance relative to its initial value and transform the chemistry into the carbon-rich regime.  Even during HBB there is enough dredge-up for our models to attain $\text{C/O} > 1$.  This additional carbon leads to familiar effects: there is an increase in opacity which increases radius and reduces the effective temperature relative to the composition-independent low-T opacity models.  These structural changes then have several consequences:
\begin{itemize}
\renewcommand{\labelitemi}{$\bullet$}
\item there is a reduction in the temperature at the base of the convective envelope;
\item acceleration of mass loss;
\item shortening of the AGB lifetime;
\item a decrease in total C+N+O yield; and
\item an increase in the lower mass limit for hot bottom burning.
\end{itemize}
The inclusion of composition-dependent low-T opacity is a necessary and feasible step towards more realistic models.  We note, however, that the degree of the consequences is given by a complex interplay between many factors such as mass loss, dredge-up efficiency, the treatment of convective overshooting, and convection theory.  In our models, the efficient dredge-up (we find $\lambda > 0.9$ in HBB models for example) and our use of the \citet{1993ApJ...413..641V} mass loss rate both contribute to its importance.

\subsection{Low-mass models}
The effect of the low-T opacity on the evolution of low-mass models depends on the extent of the third dredge-up.  It does not affect the yields of our 1.25\,M$_\odot$ models (which only have one 3DU episode), while we see a factor of ten reduction in total C+N+O yield for our 1.75\,M$_\odot$ model with composition-dependent low-T opacity (which had 12 3DU episodes compared to 92).  

\subsection{Borderline HBB models}
HBB is avoided in our $\text{[Fe/H]} = -2$ models with composition-dependent low-T opacity in a wider mass range than the $Z=0.001$ models in \citet{2010MNRAS.408.2476V}.  We were able to quantify the increase in the HBB threshold to be $\Delta M_\text{HBB} =  1.25$\,M$_\odot$, up to about 3\,M$_\odot$.  Because the dredged-up carbon is prevented from being converted into nitrogen, the opacity continues to increase and the evolution of these models diverges further.   The AGB lifetime is reduced by a factor of about three and the number of thermal pulses by a factor of 10, while the final core mass is reduced by more than 0.1\,M$_\odot$.  The difference in chemical yields is no less extreme: $Y$ is decreased by up to 0.16 and $\Delta\text{[C/Fe]}$, $\Delta\text{[N/Fe]}$, and $\Delta\text{[O/Fe]}$ are around $+1.0$, $-2.7$, and $-0.7$ respectively, where these are the defined as the yield of the models with composition-dependent low-T opacity relative to those without it.  These effects are also apparent in our 2.5\,M$_\odot$ $Z=0$ model, demonstrating that there is no metallicity limit below which composition-dependent low-T opacity may be safely neglected. 

\subsection{HBB models}
In models with composition-dependent low-T opacity that are massive enough for HBB the effect on the structure is minimal.  The use of composition-dependent low-T opacity causes a small decrease in $T_\text{bce}$ (of about 1 to 3 MK) which slightly affects the surface abundances.  The main difference emerges as a truncation of the AGB due to the faster mass loss rate.  In the models with [Fe/H] $\leq -2$ there is about a factor of 2 reduction in lifetime and a roughly concordant decrease in C, N, O, and $Y$ yields.  Since the core growth rate is not affected by the opacity treatment, the shorter lifetime reduces the final core mass (by around 0.04\,M$_\odot$ in these models).

\subsection{Implications for chemical evolution}
Composition-dependent low-T opacity reduces the oxygen and total C+N+O yield in intermediate-mass $\text{[Fe/H]} \leq -2$ models.  The degree of these two changes, however, is not strong enough to support AGB stars as being the polluters in the globular cluster self-pollution scenario since our models still do not show nett oxygen depletion and the C+N+O yield is high compared to the observed internal spreads.  The effects of composition-dependent low-T opacity may help to explain the observed high CEMP/EMP and CEMP/NEMP ratios because including it in models increases the stellar mass limit for HBB.  In binary systems this increases the potential number of donors to make CEMP stars at the expense of NEMP stars.  There are also suggestions that the prevalence of CEMP stars requires a lower stellar mass limit for the third dredge-up \citep{2009A&A...508.1359I}.  Composition-dependent low-T opacity would be crucial in models of such carbon-rich stars.

\bibliographystyle{apj}
\bibliography{apj-jour}

\begin{thebibliography}{}
\expandafter\ifx\csname natexlab\endcsname\relax\def\natexlab#1{#1}\fi

\bibitem[{{Alexander}(1975)}]{1975ApJS...29..363A}
{Alexander}, D.~R. 1975, \apjs, 29, 363

\bibitem[{{Alexander} \& {Ferguson}(1994)}]{1994ApJ...437..879A}
{Alexander}, D.~R., \& {Ferguson}, J.~W. 1994, \apj, 437, 879

\bibitem[{{Alexander} {et~al.}(1983){Alexander}, {Rypma}, \&
  {Johnson}}]{1983ApJ...272..773A}
{Alexander}, D.~R., {Rypma}, R.~L., \& {Johnson}, H.~R. 1983, \apj, 272, 773

\bibitem[{{Asplund} {et~al.}(2009){Asplund}, {Grevesse}, {Sauval}, \&
  {Scott}}]{2009ARA&A..47..481A}
{Asplund}, M., {Grevesse}, N., {Sauval}, A.~J., \& {Scott}, P. 2009, \araa, 47,
  481

\bibitem[{{Bessell} {et~al.}(1989){Bessell}, {Brett}, {Wood}, \&
  {Scholz}}]{1989A&AS...77....1B}
{Bessell}, M.~S., {Brett}, J.~M., {Wood}, P.~R., \& {Scholz}, M. 1989, \aaps,
  77, 1

\bibitem[{{Bloecker}(1995)}]{1995A&A...297..727B}
{Bloecker}, T. 1995, \aap, 297, 727

\bibitem[{{Busso} {et~al.}(1999){Busso}, {Gallino}, \&
  {Wasserburg}}]{1999ARA&A..37..239B}
{Busso}, M., {Gallino}, R., \& {Wasserburg}, G.~J. 1999, \araa, 37, 239

\bibitem[{{Campbell} \& {Lattanzio}(2008)}]{2008A&A...490..769C}
{Campbell}, S.~W., \& {Lattanzio}, J.~C. 2008, \aap, 490, 769

\bibitem[{{Carollo} {et~al.}(2012){Carollo}, {Beers}, {Bovy}, {Sivarani},
  {Norris}, {Freeman}, {Aoki}, {Lee}, \& {Kennedy}}]{2012ApJ...744..195C}
{Carollo}, D., {Beers}, T.~C., {Bovy}, J., {et~al.} 2012, \apj, 744, 195

\bibitem[{{Carretta} {et~al.}(2010){Carretta}, {Bragaglia}, {Gratton},
  {Recio-Blanco}, {Lucatello}, {D'Orazi}, \& {Cassisi}}]{2010A&A...516A..55C}
{Carretta}, E., {Bragaglia}, A., {Gratton}, R.~G., {et~al.} 2010, \aap, 516,
  A55

\bibitem[{{Cassisi} {et~al.}(1996){Cassisi}, {Castellani}, \&
  {Tornambe}}]{1996ApJ...459..298C}
{Cassisi}, S., {Castellani}, V., \& {Tornambe}, A. 1996, \apj, 459, 298

\bibitem[{{Castellani} {et~al.}(1985){Castellani}, {Chieffi}, {Tornambe}, \&
  {Pulone}}]{1985ApJ...296..204C}
{Castellani}, V., {Chieffi}, A., {Tornambe}, A., \& {Pulone}, L. 1985, \apj,
  296, 204

\bibitem[{{Cohen} \& {Mel{\'e}ndez}(2005)}]{2005AJ....129..303C}
{Cohen}, J.~G., \& {Mel{\'e}ndez}, J. 2005, \aj, 129, 303

\bibitem[{{Cottrell} \& {Da Costa}(1981)}]{1981ApJ...245L..79C}
{Cottrell}, P.~L., \& {Da Costa}, G.~S. 1981, \apjl, 245, L79

\bibitem[{{Cristallo} {et~al.}(2009){Cristallo}, {Straniero}, {Gallino},
  {Piersanti}, {Dom{\'{\i}}nguez}, \& {Lederer}}]{2009ApJ...696..797C}
{Cristallo}, S., {Straniero}, O., {Gallino}, R., {et~al.} 2009, \apj, 696, 797

\bibitem[{{Cristallo} {et~al.}(2007){Cristallo}, {Straniero}, {Lederer}, \&
  {Aringer}}]{2007ApJ...667..489C}
{Cristallo}, S., {Straniero}, O., {Lederer}, M.~T., \& {Aringer}, B. 2007,
  \apj, 667, 489

\bibitem[{{Cristallo} {et~al.}(2011){Cristallo}, {Piersanti}, {Straniero},
  {Gallino}, {Dom{\'{\i}}nguez}, {Abia}, {Di Rico}, {Quintini}, \&
  {Bisterzo}}]{2011ApJS..197...17C}
{Cristallo}, S., {Piersanti}, L., {Straniero}, O., {et~al.} 2011, \apjs, 197,
  17

\bibitem[{{D'Antona} {et~al.}(2005){D'Antona}, {Bellazzini}, {Caloi}, {Pecci},
  {Galleti}, \& {Rood}}]{2005ApJ...631..868D}
{D'Antona}, F., {Bellazzini}, M., {Caloi}, V., {et~al.} 2005, \apj, 631, 868

\bibitem[{{D'Orazi} {et~al.}(2013){D'Orazi}, {Campbell}, {Lugaro}, {Lattanzio},
  {Pignatari}, \& {Carretta}}]{2013MNRAS.433..366D}
{D'Orazi}, V., {Campbell}, S.~W., {Lugaro}, M., {et~al.} 2013, \mnras, 433, 366

\bibitem[{{Ferguson} {et~al.}(2005){Ferguson}, {Alexander}, {Allard}, {Barman},
  {Bodnarik}, {Hauschildt}, {Heffner-Wong}, \& {Tamanai}}]{2005ApJ...623..585F}
{Ferguson}, J.~W., {Alexander}, D.~R., {Allard}, F., {et~al.} 2005, \apj, 623,
  585

\bibitem[{{Fishlock} {et~al.}(2014){Fishlock}, {Karakas}, \&
  {Stancliffe}}]{2014MNRAS.tmp....3F}
{Fishlock}, C.~K., {Karakas}, A.~I., \& {Stancliffe}, R.~J. 2014, \mnras,
  doi:10.1093/mnras/stt2313

\bibitem[{{Frebel} {et~al.}(2006){Frebel}, {Christlieb}, {Norris}, {Beers},
  {Bessell}, {Rhee}, {Fechner}, {Marsteller}, {Rossi}, {Thom}, {Wisotzki}, \&
  {Reimers}}]{2006ApJ...652.1585F}
{Frebel}, A., {Christlieb}, N., {Norris}, J.~E., {et~al.} 2006, \apj, 652, 1585

\bibitem[{{Frost} \& {Lattanzio}(1996)}]{1996ApJ...473..383F}
{Frost}, C.~A., \& {Lattanzio}, J.~C. 1996, \apj, 473, 383

\bibitem[{{Gratton} {et~al.}(2004){Gratton}, {Sneden}, \&
  {Carretta}}]{2004ARA&A..42..385G}
{Gratton}, R., {Sneden}, C., \& {Carretta}, E. 2004, \araa, 42, 385

\bibitem[{{Gratton} {et~al.}(2012){Gratton}, {Lucatello}, {Carretta},
  {Bragaglia}, {D'Orazi}, {Al Momany}, {Sollima}, {Salaris}, \&
  {Cassisi}}]{2012A&A...539A..19G}
{Gratton}, R.~G., {Lucatello}, S., {Carretta}, E., {et~al.} 2012, \aap, 539,
  A19

\bibitem[{{Grevesse} {et~al.}(2007){Grevesse}, {Asplund}, \&
  {Sauval}}]{2007SSRv..130..105G}
{Grevesse}, N., {Asplund}, M., \& {Sauval}, A.~J. 2007, \ssr, 130, 105

\bibitem[{{Herwig}(2000)}]{2000A&A...360..952H}
{Herwig}, F. 2000, \aap, 360, 952

\bibitem[{{Ivans} {et~al.}(1999){Ivans}, {Sneden}, {Kraft}, {Suntzeff},
  {Smith}, {Langer}, \& {Fulbright}}]{1999AJ....118.1273I}
{Ivans}, I.~I., {Sneden}, C., {Kraft}, R.~P., {et~al.} 1999, \aj, 118, 1273

\bibitem[{{Iwamoto}(2009)}]{2009PASA...26..145I}
{Iwamoto}, N. 2009, \pasa, 26, 145

\bibitem[{{Izzard} {et~al.}(2009){Izzard}, {Glebbeek}, {Stancliffe}, \&
  {Pols}}]{2009A&A...508.1359I}
{Izzard}, R.~G., {Glebbeek}, E., {Stancliffe}, R.~J., \& {Pols}, O.~R. 2009,
  \aap, 508, 1359

\bibitem[{{Johnson} {et~al.}(2007){Johnson}, {Herwig}, {Beers}, \&
  {Christlieb}}]{2007ApJ...658.1203J}
{Johnson}, J.~A., {Herwig}, F., {Beers}, T.~C., \& {Christlieb}, N. 2007, \apj,
  658, 1203

\bibitem[{{Karakas} {et~al.}(2002){Karakas}, {Lattanzio}, \&
  {Pols}}]{2002PASA...19..515K}
{Karakas}, A.~I., {Lattanzio}, J.~C., \& {Pols}, O.~R. 2002, \pasa, 19, 515

\bibitem[{{Kennedy} {et~al.}(2011){Kennedy}, {Sivarani}, {Beers}, {Lee},
  {Placco}, {Rossi}, {Christlieb}, {Herwig}, \& {Plez}}]{2011AJ....141..102K}
{Kennedy}, C.~R., {Sivarani}, T., {Beers}, T.~C., {et~al.} 2011, \aj, 141, 102

\bibitem[{{Kitsikis} \& {Weiss}(2008)}]{2008ASPC..388..183K}
{Kitsikis}, A., \& {Weiss}, A. 2008, in Astronomical Society of the Pacific
  Conference Series, Vol. 388, Mass Loss from Stars and the Evolution of
  Stellar Clusters, ed. A.~{de Koter}, L.~J. {Smith}, \& L.~B.~F.~M. {Waters},
  183

\bibitem[{{Lattanzio}(1986)}]{1986ApJ...311..708L}
{Lattanzio}, J.~C. 1986, \apj, 311, 708

\bibitem[{{Lederer} \& {Aringer}(2009)}]{2009A&A...494..403L}
{Lederer}, M.~T., \& {Aringer}, B. 2009, \aap, 494, 403

\bibitem[{{Lee} {et~al.}(2005){Lee}, {Joo}, {Han}, {Chung}, {Ree}, {Sohn},
  {Kim}, {Yoon}, {Yi}, \& {Demarque}}]{2005ApJ...621L..57L}
{Lee}, Y.-W., {Joo}, S.-J., {Han}, S.-I., {et~al.} 2005, \apjl, 621, L57

\bibitem[{{Lucatello} {et~al.}(2006){Lucatello}, {Beers}, {Christlieb},
  {Barklem}, {Rossi}, {Marsteller}, {Sivarani}, \& {Lee}}]{2006ApJ...652L..37L}
{Lucatello}, S., {Beers}, T.~C., {Christlieb}, N., {et~al.} 2006, \apjl, 652,
  L37

\bibitem[{{Lucatello} {et~al.}(2005){Lucatello}, {Tsangarides}, {Beers},
  {Carretta}, {Gratton}, \& {Ryan}}]{2005ApJ...625..825L}
{Lucatello}, S., {Tsangarides}, S., {Beers}, T.~C., {et~al.} 2005, \apj, 625,
  825

\bibitem[{{Lucy} {et~al.}(1986){Lucy}, {Robertson}, \&
  {Sharp}}]{1986A&A...154..267L}
{Lucy}, L.~B., {Robertson}, J.~A., \& {Sharp}, C.~M. 1986, \aap, 154, 267

\bibitem[{{Lugaro} {et~al.}(2012){Lugaro}, {Karakas}, {Stancliffe}, \&
  {Rijs}}]{2012ApJ...747....2L}
{Lugaro}, M., {Karakas}, A.~I., {Stancliffe}, R.~J., \& {Rijs}, C. 2012, \apj,
  747, 2

\bibitem[{{Marigo}(2002)}]{2002A&A...387..507M}
{Marigo}, P. 2002, \aap, 387, 507

\bibitem[{{Marigo}(2007)}]{2007A&A...467.1139M}
---. 2007, \aap, 467, 1139

\bibitem[{{Marigo} \& {Aringer}(2009)}]{2009A&A...508.1539M}
{Marigo}, P., \& {Aringer}, B. 2009, \aap, 508, 1539

\bibitem[{{Marigo} {et~al.}(2013){Marigo}, {Bressan}, {Nanni}, {Girardi}, \&
  {Pumo}}]{2013MNRAS.434..488M}
{Marigo}, P., {Bressan}, A., {Nanni}, A., {Girardi}, L., \& {Pumo}, M.~L. 2013,
  \mnras, 434, 488

\bibitem[{{Masseron} {et~al.}(2010){Masseron}, {Johnson}, {Plez}, {van Eck},
  {Primas}, {Goriely}, \& {Jorissen}}]{2010A&A...509A..93M}
{Masseron}, T., {Johnson}, J.~A., {Plez}, B., {et~al.} 2010, \aap, 509, A93

\bibitem[{{McSaveney} {et~al.}(2007){McSaveney}, {Wood}, {Scholz}, {Lattanzio},
  \& {Hinkle}}]{2007MNRAS.378.1089M}
{McSaveney}, J.~A., {Wood}, P.~R., {Scholz}, M., {Lattanzio}, J.~C., \&
  {Hinkle}, K.~H. 2007, \mnras, 378, 1089

\bibitem[{{Milone} {et~al.}(2013){Milone}, {Marino}, {Piotto}, {Bedin},
  {Anderson}, {Aparicio}, {Bellini}, {Cassisi}, {D'Antona}, {Grundahl},
  {Monelli}, \& {Yong}}]{2013ApJ...767..120M}
{Milone}, A.~P., {Marino}, A.~F., {Piotto}, G., {et~al.} 2013, \apj, 767, 120

\bibitem[{{Nieuwenhuijzen} \& {de Jager}(1990)}]{1990A&A...231..134N}
{Nieuwenhuijzen}, H., \& {de Jager}, C. 1990, \aap, 231, 134

\bibitem[{{Paczy{\'n}ski}(1970)}]{1970AcA....20...47P}
{Paczy{\'n}ski}, B. 1970, \actaa, 20, 47

\bibitem[{{Pasquini} {et~al.}(2011){Pasquini}, {Mauas}, {K{\"a}ufl}, \&
  {Cacciari}}]{2011A&A...531A..35P}
{Pasquini}, L., {Mauas}, P., {K{\"a}ufl}, H.~U., \& {Cacciari}, C. 2011, \aap,
  531, A35

\bibitem[{{Paxton} {et~al.}(2013){Paxton}, {Cantiello}, {Arras}, {Bildsten},
  {Brown}, {Dotter}, {Mankovich}, {Montgomery}, {Stello}, {Timmes}, \&
  {Townsend}}]{2013ApJS..208....4P}
{Paxton}, B., {Cantiello}, M., {Arras}, P., {et~al.} 2013, \apjs, 208, 4

\bibitem[{{Piotto} {et~al.}(2007){Piotto}, {Bedin}, {Anderson}, {King},
  {Cassisi}, {Milone}, {Villanova}, {Pietrinferni}, \&
  {Renzini}}]{2007ApJ...661L..53P}
{Piotto}, G., {Bedin}, L.~R., {Anderson}, J., {et~al.} 2007, \apjl, 661, L53

\bibitem[{{Pols} {et~al.}(2012){Pols}, {Izzard}, {Stancliffe}, \&
  {Glebbeek}}]{2012A&A...547A..76P}
{Pols}, O.~R., {Izzard}, R.~G., {Stancliffe}, R.~J., \& {Glebbeek}, E. 2012,
  \aap, 547, A76

\bibitem[{{Reimers}(1975)}]{1975MSRSL...8..369R}
{Reimers}, D. 1975, Memoires of the Societe Royale des Sciences de Liege, 8,
  369

\bibitem[{{Rogers} \& {Nayfonov}(2002)}]{2002ApJ...576.1064R}
{Rogers}, F.~J., \& {Nayfonov}, A. 2002, \apj, 576, 1064

\bibitem[{{Rossi} {et~al.}(2005){Rossi}, {Beers}, {Sneden}, {Sevastyanenko},
  {Rhee}, \& {Marsteller}}]{2005AJ....130.2804R}
{Rossi}, S., {Beers}, T.~C., {Sneden}, C., {et~al.} 2005, \aj, 130, 2804

\bibitem[{{Sackmann} \& {Boothroyd}(1991)}]{1991ApJ...366..529S}
{Sackmann}, I.-J., \& {Boothroyd}, A.~I. 1991, \apj, 366, 529

\bibitem[{{Scalo} {et~al.}(1975){Scalo}, {Despain}, \&
  {Ulrich}}]{1975ApJ...196..805S}
{Scalo}, J.~M., {Despain}, K.~H., \& {Ulrich}, R.~K. 1975, \apj, 196, 805

\bibitem[{{Scalo} \& {Ulrich}(1975)}]{1975ApJ...200..682S}
{Scalo}, J.~M., \& {Ulrich}, R.~K. 1975, \apj, 200, 682

\bibitem[{{Siess} {et~al.}(2002){Siess}, {Livio}, \&
  {Lattanzio}}]{2002ApJ...570..329S}
{Siess}, L., {Livio}, M., \& {Lattanzio}, J. 2002, \apj, 570, 329

\bibitem[{{Spite} {et~al.}(2005){Spite}, {Cayrel}, {Plez}, {Hill}, {Spite},
  {Depagne}, {Fran{\c c}ois}, {Bonifacio}, {Barbuy}, {Beers}, {Andersen},
  {Molaro}, {Nordstr{\"o}m}, \& {Primas}}]{2005A&A...430..655S}
{Spite}, M., {Cayrel}, R., {Plez}, B., {et~al.} 2005, \aap, 430, 655

\bibitem[{{Stancliffe} \& {Glebbeek}(2008)}]{2008MNRAS.389.1828S}
{Stancliffe}, R.~J., \& {Glebbeek}, E. 2008, \mnras, 389, 1828

\bibitem[{{Stancliffe} {et~al.}(2007){Stancliffe}, {Glebbeek}, {Izzard}, \&
  {Pols}}]{2007A&A...464L..57S}
{Stancliffe}, R.~J., {Glebbeek}, E., {Izzard}, R.~G., \& {Pols}, O.~R. 2007,
  \aap, 464, L57

\bibitem[{{Stancliffe} {et~al.}(2004){Stancliffe}, {Tout}, \&
  {Pols}}]{2004MNRAS.352..984S}
{Stancliffe}, R.~J., {Tout}, C.~A., \& {Pols}, O.~R. 2004, \mnras, 352, 984

\bibitem[{{Straniero} {et~al.}(2006){Straniero}, {Gallino}, \&
  {Cristallo}}]{2006NuPhA.777..311S}
{Straniero}, O., {Gallino}, R., \& {Cristallo}, S. 2006, Nuclear Physics A,
  777, 311

\bibitem[{{Suda} \& {Fujimoto}(2010)}]{2010MNRAS.405..177S}
{Suda}, T., \& {Fujimoto}, M.~Y. 2010, \mnras, 405, 177

\bibitem[{{Suda} {et~al.}(2011){Suda}, {Yamada}, {Katsuta}, {Komiya},
  {Ishizuka}, {Aoki}, \& {Fujimoto}}]{2011MNRAS.412..843S}
{Suda}, T., {Yamada}, S., {Katsuta}, Y., {et~al.} 2011, \mnras, 412, 843

\bibitem[{{Suda} {et~al.}(2013){Suda}, {Komiya}, {Yamada}, {Katsuta}, {Aoki},
  {Gil-Pons}, {Doherty}, {Campbell}, {Wood}, \&
  {Fujimoto}}]{2013MNRAS.432L..46S}
{Suda}, T., {Komiya}, Y., {Yamada}, S., {et~al.} 2013, \mnras, 432, L46

\bibitem[{{Timmes} \& {Arnett}(1999)}]{1999ApJS..125..277T}
{Timmes}, F.~X., \& {Arnett}, D. 1999, \apjs, 125, 277

\bibitem[{{Timmes} \& {Swesty}(2000)}]{2000ApJS..126..501T}
{Timmes}, F.~X., \& {Swesty}, F.~D. 2000, \apjs, 126, 501

\bibitem[{{Vassiliadis} \& {Wood}(1993)}]{1993ApJ...413..641V}
{Vassiliadis}, E., \& {Wood}, P.~R. 1993, \apj, 413, 641

\bibitem[{{Ventura} {et~al.}(2009){Ventura}, {Caloi}, {D'Antona}, {Ferguson},
  {Milone}, \& {Piotto}}]{2009MNRAS.399..934V}
{Ventura}, P., {Caloi}, V., {D'Antona}, F., {et~al.} 2009, \mnras, 399, 934

\bibitem[{{Ventura} {et~al.}(2013){Ventura}, {Criscienzo}, {D'Antona},
  {Vesperini}, {Tailo}, {Dell'Agli}, \& {D'Ercole}}]{2013MNRAS.tmp.2760V}
{Ventura}, P., {Criscienzo}, M.~D., {D'Antona}, F., {et~al.} 2013, \mnras,
  arXiv:1311.0683

\bibitem[{{Ventura} \& {Marigo}(2009)}]{2009MNRAS.399L..54V}
{Ventura}, P., \& {Marigo}, P. 2009, \mnras, 399, L54

\bibitem[{{Ventura} \& {Marigo}(2010)}]{2010MNRAS.408.2476V}
---. 2010, \mnras, 408, 2476

\bibitem[{{Villanova} {et~al.}(2010){Villanova}, {Geisler}, \&
  {Piotto}}]{2010ApJ...722L..18V}
{Villanova}, S., {Geisler}, D., \& {Piotto}, G. 2010, \apjl, 722, L18

\bibitem[{{Weiss} \& {Ferguson}(2009)}]{2009A&A...508.1343W}
{Weiss}, A., \& {Ferguson}, J.~W. 2009, \aap, 508, 1343

\bibitem[{Wood(1990)}]{Wood1989}
Wood, P.~R. 1990, From Miras to Planetary Nebulae: Which Path to Stellar
  Evolution? : Montpellier, France, September 4-7, 1989, IAP astrophysics
  meeting (Editions Fronti{\`e}res)

\bibitem[{{Yong} {et~al.}(2009){Yong}, {Grundahl}, {D'Antona}, {Karakas},
  {Lattanzio}, \& {Norris}}]{2009ApJ...695L..62Y}
{Yong}, D., {Grundahl}, F., {D'Antona}, F., {et~al.} 2009, \apjl, 695, L62

\end{thebibliography}

\end{document}